\documentclass[aps,prd,amsmath,floats,floatfix,twocolumn,
  superscriptaddress,nofootinbib,showpacs,10pt,longbibliography]{revtex4-2}

\pdfoutput=1 
\usepackage{diagbox}
\usepackage{multirow}
\usepackage{braket}
\usepackage{amsfonts}
\usepackage{amsmath}
\usepackage{amssymb}
\usepackage{mathtools}
\usepackage{amsthm}
\usepackage{bm}
\usepackage{dcolumn}
\usepackage{epsfig}
\usepackage{commath}
\usepackage{graphicx}
\usepackage{graphics}
\usepackage[latin1]{inputenc}
\usepackage{latexsym}
\usepackage{rotating}
\usepackage[dvipsnames]{xcolor}
\usepackage{mathrsfs}
\usepackage{microtype}
\usepackage{verbatim}
\usepackage{url}

\usepackage[caption=false]{subfig}
\usepackage[normalem]{ulem}
\usepackage{tikz}

\usepackage[breaklinks=true]{hyperref}
\hypersetup{
    colorlinks=true,
    linkcolor=NavyBlue,
    filecolor=Magenta,
    urlcolor=NavyBlue,
    citecolor=NavyBlue
}

\usepackage{yfonts}
\usepackage{float}
\usepackage{xspace} 
\usepackage{mathrsfs}
\usepackage[toc,page]{appendix}
\usepackage{siunitx}
\usepackage{array}
\usepackage[normalem]{ulem}

\makeatletter
\newcommand*{\rom}[1]{\expandafter\@slowromancap\romannumeral #1@}
\makeatother
\allowdisplaybreaks

\AtBeginDocument{%
    \newwrite\bibnotes
    \def\bibnotesext{Notes.bib}
    \immediate\openout\bibnotes=\jobname\bibnotesext
    \immediate\write\bibnotes{@CONTROL{REVTEX41Control}}
    \immediate\write\bibnotes{@CONTROL{%
    apsrev41Control,author="08",editor="1",pages="1",title="0",year="1"}}
    \if@filesw
    \immediate\write\@auxout{\string\citation{apsrev41Control}}%
    \fi
}

\usepackage{stackengine}

\newcommand{\AEI}{Max Planck Institute for Gravitational Physics (Albert Einstein Institute), D-14476 Potsdam, Germany}

\begin{document}

\title{Beyond Three Terms: \\ Continued Fractions for Rotating Black Holes in Modified Gravity}
\author{Georgios Karikos}
\email{karikos2@illinois.edu}
\affiliation{Illinois Center for Advanced Studies of the Universe, Department of Physics, University of Illinois Urbana-Champaign, Urbana, IL 61801, USA}
\author{Jayana A. Saes}
\email{jayanaa2@illinois.edu}
\affiliation{Illinois Center for Advanced Studies of the Universe, Department of Physics, University of Illinois Urbana-Champaign, Urbana, IL 61801, USA}
\author{Pratik Wagle}
\email{pratik.wagle@aei.mpg.de}
\affiliation{\AEI}
\author{Nicol\'as Yunes}
\email{nyunes@illinois.edu}
\affiliation{Illinois Center for Advanced Studies of the Universe, Department of Physics, University of Illinois Urbana-Champaign, Urbana, IL 61801, USA}

\date{\today}

\begin{abstract}

Black-hole ringdown offers a clean probe of strong gravity, but one of its most accurate tools--Leaver's continued-fraction method--requires a three-term recurrence relation. Beyond general relativity, and more generally in non-Kerr spacetimes, Frobenius expansions of the perturbation equations generically produce higher-order recurrence relations and, often, couplings among the series coefficients, obstructing a direct application of Leaver's method. Here we develop a general reduction scheme that maps arbitrary scalar and matrix $N$-term recurrence relations to a three-term form, thereby extending continued fractions to a broad class of perturbation problems in modified gravity. As an application, we compute the quasinormal-mode spectrum of slowly-rotating black holes in dynamical Chern-Simons gravity, where the polar sector yields a 16-term, decoupled, scalar recurrence relation, and the axial sector yields a 12-term, coupled, matrix recurrence relation. After applying our reduction scheme, both systems can be solved with continued fractions. For the fundamental $(\ell,m)=(2,2)$ mode, our results agree well with independent calculations based on eigenvalue-perturbation and metric/spectral methods across the parameter range studied. This framework provides a robust and practical route to precision ringdown calculations beyond the standard three-term setting and supports tests of gravity with current and future gravitational-wave observations.

\end{abstract}
\maketitle
\allowdisplaybreaks[4] 
\section{Introduction}

The direct detection of gravitational waves, beginning with the GW150914 event and followed by the growing catalog of the LIGO-Virgo-KAGRA (LVK) network, opened an observational window into the strong, dynamical regime of gravity \cite{LIGOScientific:2016aoc,LIGOScientific:2021sio}. Unlike Solar System tests, which have validated general relativity (GR) primarily in weak-field settings \cite{Will:2014kxa}, compact-binary coalescences let us confront GR where gravity is nonlinear and directly dynamical~\cite{Yunes:2025xwp}. This is no longer a purely forward-looking program. Gravitational-wave observations already enable a growing suite of tests of GR~\cite{Yunes:2016jcc,Nair:2019iur,Perkins:2021mhb,Lagos:2024boe,Xie:2025voe,Yagi:2016jml,Cardoso:2016ryw,Isi:2019aib,Islam:2021pbd,Capano:2021etf,Isi:2021iql,Wagle:2019mdq,Schumacher:2023jxq}, including increasingly quantitative studies of black-hole spectroscopy \cite{LIGOScientific:2016lio,LIGOScientific:2021sio,Meidam:2014jpa,Bhagwat:2019dtm,Ota:2021ypb,CalderonBustillo:2020rmh,Berti:2009kk,Silva:2022srr}. With continued sensitivity improvements and the advent of next-generation detectors, these tests can only become sharper \cite{Yunes:2025xwp,Gupta:2025utd}.

The ringdown stage of compact-binary coalescence is one of the cleanest probes of strong gravity. Once the nonlinear merger has subsided, the remnant settles toward a stationary black hole, and the late-time waveform is described by a discrete set of exponentially damped sinusoids, called quasinormal modes (QNMs). In GR, that spectrum is fixed by the mass and spin of the remnant Kerr state, so measuring multiple modes directly tests both the Kerr nature of the remnant and the no-hair structure of the solution \cite{Teukolsky:1973ha,Press:1973zz,Leaver:1985ax,Berti:2009kk}. Because the ringdown isolates the linear response of the final object from the full nonlinear complexity of the merger, it offers a particularly sharp way to search for new gravitational dynamics. Any robust inconsistency among the observed modes, or any statistically significant shift away from the Kerr spectrum, would point to new degrees of freedom or modified field equations.

This is precisely why ringdown is so important beyond GR. Many extensions of Einstein's theory introduce additional fields, higher-curvature operators, or parity-violating interactions, all of which can modify black-hole dynamics and leave direct imprints in the QNM spectrum \cite{Berti:2015itd,Alexander:2009tp,Berti:2009kk,Maselli:2023khq}. Gravitational-wave observations, and ringdown in particular, therefore provide a direct route to constrain such theories in the regime where deviations from Kerr should be most cleanly exposed \cite{Berti:2018cxi,Wagle:2021tam,Li:2022pcy,Wagle:2023fwl,Li:2025fci,Chung:2025gyg,Hussain:2022ins,Molina:2010fb,Blazquez-Salcedo:2016enn,Cano:2023jbk,Cano:2023tmv,Cano:2024bhh,Cano:2024ezp,Cano:2024jkd,Srivastava:2021imr,Alapati:2025jmd,Chung:2025wbg,Husken:2026axq,Chung:2024ira,Chung:2024vaf,Pierini:2021jxd,Pierini:2022eim}.

The theoretical framework for this problem is black-hole perturbation theory, where one studies perturbations of a stationary background spacetime \cite{Maggiore:2018sht,Chandrasekhar:1985kt}. Within GR, the linear metric perturbations obey the Regge-Wheeler and Zerilli-Moncrief equations in a Schwarzschild background \cite{Regge:1957td,Zerilli:1970se,Moncrief:1974am}, while the linear curvature perturbations obey the Teukolsky equation in a Kerr background \cite{Teukolsky:1973ha}. Outside GR, the metric and curvature perturbation formalisms have to be extended to account for new radiative degrees of freedom and their couplings to tensor perturbations \cite{Teukolsky:1973ha,Li:2022pcy,Hussain:2022ins,Cano:2023tmv,Chung:2024ira}. Once the ``master'' perturbation equations have been identified, the central task is to compute the QNM spectrum with sufficient accuracy that ringdown can be used as a precision test of gravity, rather than merely a qualitative one. 

This task is nontrivial because QNMs are defined by a boundary-value problem with complex frequencies: one must impose purely-ingoing waves at the horizon and purely-outgoing waves at spatial infinity. Several numerical strategies have been developed to solve this problem, including shooting methods and spectral collocation techniques \cite{Berti:2009kk,Chung:2023wkd,Chung:2023zdq}. These approaches are useful and widely employed, but they can suffer from numerical contamination, limited stability, or approximations tied to the discretization scheme. By contrast, Leaver's continued-fraction method \cite{Leaver:1985ax} is especially powerful when it applies. In that approach, one expands the radial solution in a Frobenius series about the horizon (or another regular singular point), obtains a three-term recurrence relation for the series coefficients, and determines the QNM frequencies by imposing convergence of the associated continued fraction. The result is a method with excellent numerical stability and very high accuracy, which is precisely why continued-fraction methods have become a benchmark tool for QNM calculations~\cite{Pani:2013wsa,Roussille:2023sdr,Deng:2025atg}.

The difficulty in applying the continued-fraction method beyond GR is easy to state: Leaver's construction requires a scalar three-term recurrence relation, while beyond-GR perturbation equations are generically much more complicated. In particular, two distinct obstructions arise when trying to apply the continued-fraction method beyond GR. The first is structural. The Frobenius expansion typically yields an $(N>3)$-term recurrence relation. The second is dynamical. The perturbation equations are coupled, so the series coefficients are vector-valued and satisfy a matrix recurrence relation. Either feature prevents the direct application of the standard continued-fraction method, and in modified gravity, they often occur \textit{together}. 

Previous work made important progress on each of these obstructions separately, but only in restricted settings. The first challenge, namely recurrence relations with more than three terms, was addressed through reduction schemes that map higher-order relations to three-term ones, first for $4 \rightarrow 3$ reductions in~\cite{Leaver:1990zz} and later for $5\rightarrow 3$ reductions in~\cite{Onozawa:1995vu}. The second challenge, namely coupled recurrence relations, was tackled using matrix-valued continued-fraction methods in~\cite{Pani:2013wsa,Rosa:2011my,Macedo:2013jja}, with subsequent extensions in~\cite{Pani:2013pma}. However, these developments do not combine in a straightforward way. In particular, when higher-order structure and coupling are both present, as is often the case for perturbations of spinning black holes in theories beyond GR, the existing reduction and matrix-valued continued-fraction methods cannot be directly applied in tandem as the problem poses significant arithmetic and computational challenges.

In this work, we develop a generalized continued-fraction framework that overcomes both obstructions simultaneously. Our method provides a systematic reduction scheme that maps arbitrary high-order scalar and matrix recurrence relations to three-term form, thereby extending Leaver's method to a broad class of beyond-GR perturbation problems. As a concrete test case, we apply this formalism to slowly-rotating black holes in dynamical Chern--Simons (dCS) gravity, which introduces a non-minimal coupling between a (pseudo)scalar field and the metric~\cite{Jackiw:2003pm,Alexander:2009tp}. In this theory, the polar perturbation sector yields a 16-term, decoupled, scalar recurrence relation, while the axial perturbation sector yields a 12-term, coupled, matrix recurrence relation. We show here that, after implementing our reduction method, both systems can be solved with continued fractions. We benchmark our continued-fraction results for the fundamental $(2,2)$ mode against independent results obtained using the modified Teukolsky formalism~\cite{Li:2022pcy,Hussain:2022ins,Cano:2023tmv} with the eigenvalue perturbation method (EVP)~\cite{Wagle:2023fwl,Li:2025fci} and the MEtric-perTuRbatIon Spectral framework (METRICS)~\cite{Chung:2025gyg}, finding excellent agreement in both sectors. We also tabulate the remaining $\ell=2$ gravitational-led modes and the first overtone in Appendix~\ref{appendix:Appendix_A}, providing separate data files for easy access to these modes~\cite{github_repo}. More broadly, this establishes the continued-fraction method as a practical route to Leaver-type QNM calculations for slowly-rotating black holes beyond GR.

The remainder of this paper presents the details of the calculations described above, and it is organized as follows. In Sec.~\ref{sec:Review}, we review the continued-fraction method for both single ordinary differential equations (ODEs) in the context of a non-rotating black hole in GR and, generically, for coupled ODEs. Section~\ref{sec:Reductionsingle} describes in detail the reduction scheme of a generic $N$-term recurrence relation to a three-term recurrence relation for decoupled recurrence relations, and Section~\ref{sec:Reductioncoupled} generalizes the method to coupled recurrence relations. We apply both methods to obtain the QNMs of perturbed slowly-rotating black holes in dCS gravity in Sec.~\ref{sec:dCS}, and we compare our results to previous literature. Our conclusions are summarized in Sec.~\ref{sec:conclusion}. Henceforth, we use the following conventions unless stated otherwise: we work in 4-dimensions with metric signature $(-,+,+,+)$~\cite{Misner:1973prb}; we use geometric units, such that $G = 1 = M$; and we set $M=1$ when plotting and tabulating the QNM frequencies.

\section{Review of the Continuous Fraction Method}
\label{sec:Review}

In this section, we briefly review Leaver's continued-fraction method for three-term recurrence relations~\cite{Leaver:1985ax}, which is used to compute the QNMs of non-rotating black holes in GR from the Regge-Wheeler and Zerilli-Moncrief equations. We then review the corresponding generalization to coupled systems that are governed by three-term matrix recurrence relations.

\subsection{Decoupled Ordinary Differential Equations}
\label{sec:rwleaver}

The continued-fraction method for solving ODEs has been widely applied to a variety of eigenvalue problems in GR. Applying this method requires that one recast the ODE in terms of a three-term recurrence relation, which is achieved through a Frobenius series ansatz (see Sec.~\ref{sec:Reductionsingle}). In the GR context, the continued-fraction method was first proposed by Leaver~\cite{Leaver:1985ax}, who used it to obtain the QNM frequencies of perturbed Schwarzschild and Kerr black holes. In both cases, the resulting recurrence relations have three terms, and the method can be applied directly.

As an illustrative example, we briefly review the continued-fraction method from \cite{Leaver:1985ax} for polar perturbations of a non-rotating black hole in GR. We first Fourier transform the master equation and expand the master function in spherical harmonics. The master equation then reduces to a single ODE for the radial part of the master function  $\psi_{\ell}(r,\omega)$, where $r$ is the radial coordinate, $\omega$ is the frequency, and $\ell$ is the spherical-harmonic index. This ODE, also known as the Regge-Wheeler equation, is given by

\begin{align} \label{eq:rwgr}
f(r)^2 \psi_\ell''(r)
&+ \frac{2M}{r^2} f(r) \psi_\ell'(r) \nonumber \\ 
&+ \left[\omega^2 - f(r)\left(\frac{\ell(\ell+1)}{r^2} - \frac{6M}{r^3}\right)\right]\psi_\ell(r) = 0\, ,
\end{align}
where $f(r)=1-2M/r$ is the Schwarzschild factor, and primes denote derivatives with respect to $r$.

Equation~\eqref{eq:rwgr} describes the radial evolution of Schwarzschild perturbations, so the QNM spectrum is obtained by imposing physically-relevant boundary conditions. In particular, one must require the solution to represent a purely ingoing wave at the event horizon and a purely outgoing wave at spatial infinity. Assuming the Fourier convention $e^{-i\omega t}$, these conditions read
\begin{equation}
\label{eq:BoundaryConditions1}
\psi_l (r) = \begin{cases}
    e^{-i\omega r_*}, & r\rightarrow r_{\rm H} \\
    e^{i\omega r_*}, & r\rightarrow\infty \,,
\end{cases}
\end{equation}
where $r_{\rm H}=2M$ is the horizon radius and $r_*$ is the tortoise coordinate, defined by
\begin{equation}
\label{eq:rtor}
r_* = r + 2M \ln\left|\frac{r}{2M} - 1\right| \, .
\end{equation}

The boundary conditions motivate the Frobenius ansatz
\begin{equation} \label{eq:ansatzgr}
\psi_l(r) = A(r) \sum_{n=0}^\infty x_n f(r)^n \,,
\end{equation}
where the asymptotic factor $A(r)$ enforces the QNM behavior at both boundaries. A convenient choice is
\begin{equation}
A(r)=(r-2M)^{-2 i \omega M } r^{4 i \omega M} e^{i \omega r} \,,
\end{equation}
which reproduces the ingoing wave behavior at the horizon and the outgoing wave behavior at spatial infinity. The Frobenius series connects the solution between the two asymptotic regions, with series coefficients $\{x_n\}$ as unknowns. 

Substituting this Frobenius ansatz into the ODE and collecting powers of $f(r)$ gives a recurrence relation for $n \geq 1$,
\begin{equation} \label{eq:3termGR}
\gamma_{-1}(n) x_{n-1} + \gamma_0(n) x_n + \gamma_1(n) x_{n+1} = 0 \,.
\end{equation}
This is clearly a three-term recursion relation, as it relates $x_{n-1}$, $x_n$, and $x_{n+1}$. 
The coefficients $\gamma_i(n)$ depend on $n$ and $\omega$ as follows:
\begin{align}
\gamma_1(n) &= n^2 + n(2 - 4 i M\omega) - 4 i M\omega + 1 \, , \\
\gamma_0(n) &= -l(l+1) - 2 n^2 + n(-2 + 16 iM \omega) \nonumber \\
&\quad + 32 M^2\omega^2 + 8 i M \omega  + 3 \, , \\
\gamma_{-1}(n) &= n^2 - 8 i n M \omega - 4(4 M^2 \omega^2 + 1) \, .
\end{align}
At the beginning of the series, the Frobenius coefficients are defined only for $n\geq 0$, so one may equivalently take $x_k=0$ for $k<0$. Evaluating Eq.~\eqref{eq:3termGR} at $n=0$ therefore gives

\begin{equation}
\label{eq:n0RR}
\gamma_0(0) x_0 + \gamma_1(0) x_1 = 0 \, .
\end{equation}

To apply the continued-fraction method, one introduces the ratio of successive coefficients, $r_n:= x_{n+1}/x_n$. We divide the three-term recurrence relation of Eq.~\eqref{eq:3termGR} by $x_n$. Expressing the result in terms of $r_n$, one finds
\begin{equation}
\gamma_{-1}(n)\frac{1}{r_{n-1}} + \gamma_0(n) + \gamma_1(n) r_n = 0 \, .
\end{equation}
which can be solved recursively for $r_{n-1}$ to obtain
\begin{equation}
\label{eq:Rnrecursion}
r_{n-1} = -\frac{\gamma_{-1}(n)}{\gamma_0(n) + \gamma_1(n) r_n} \, .
\end{equation}
Iterating Eq.~\eqref{eq:Rnrecursion} generates a continued-fraction representation for $r_0$, namely
\begin{equation}
\label{eq:cont-frac-single}
r_0 = -\frac{\gamma_{-1}(1)}{\gamma_0(1) - \dfrac{\gamma_1(1)\gamma_{-1}(2)}{\gamma_0(2) - \dfrac{\gamma_1(2)\gamma_{-1}(3)}{\gamma_0(3) - \cdots}}} \, .
\end{equation}

The initial condition at $n=0$ given in Eq.~\eqref{eq:n0RR} provides an independent relation between $x_0$ and $x_1$. This equation can be written in terms of $r_0$ as
\begin{equation}
\gamma_0(0) + \gamma_1(0) r_0 = 0 \, .
\end{equation}
Substituting the continued-fraction expression for $r_0$ into the above equation yields 
\begin{equation}
\label{eq:eigenvalue}
 0 = \gamma_0(0) - \gamma_1(0) \frac{\gamma_{-1}(1)}{\gamma_0(1) - \dfrac{\gamma_1(1)\gamma_{-1}(2)}{\gamma_0(2) - \dfrac{\gamma_1(2)\gamma_{-1}(3)}{\gamma_0(3) - \cdots}}}\,, 
\end{equation}
a single algebraic equation for the frequency $\omega$. The QNM frequencies of the black hole are the values of $\omega$ for which the continued fraction solution of Eq.~\eqref{eq:cont-frac-single} converges and the algebraic condition is satisfied.

Since the Frobenius series expansion for the chosen ansatz in Eq.~\eqref{eq:ansatzgr} is infinite, the resulting continued fraction is also infinite. Therefore, a numerical implementation requires truncating the sum of Eq.~\eqref{eq:ansatzgr} at a finite order $n=n_{\rm max}$. At that truncation point, one prescribes a terminal value for the ratio $r_{n_{\rm max}-1}$. We set $\gamma_i(n_{\rm max}+1)=0$, resulting in 
\begin{equation}
r_{n_{\rm max}-1} = -\frac{\gamma_{-1}(n_{\rm max})}{\gamma_{0}(n_{\rm max})} \,.
\end{equation}

This condition amounts to terminating the continued fraction in Eq.~\eqref{eq:eigenvalue} at a finite order by setting $\gamma_1(n_{\rm max}) = 0$. To ensure numerical precision, one solves for $\omega$ for larger and larger values of $n_{\rm max}$ until the solution stabilizes. Using this method, one, for example, easily finds that, for $\ell=2$ Schwarzschild gravitational perturbations in GR, the fundamental QNM frequency is
\begin{equation} \label{eq:qnmschwgr}
M\omega_{220} \approx 0.37367 - 0.08896\, i \, ,
\end{equation}
which serves as a standard benchmark for validating the implementation of the continued-fraction method. To get a precision of 5 digits, we only need a maximum of $n_{max}=14$ terms in the continued fraction; increasing to $n_{max}=15$ changes the frequency after the 6th digit. Eventually, we observe that the frequency stabilizes for $n_{max} \geq 50$ terms, if one requires 10th digit precisiom.

\subsection{Coupled Ordinary Differential Equations}

The continued-fraction method is also a powerful tool when the problem consists of a set of coupled ODEs. This version was first introduced to compute the propagation of a massive vector field on a Schwarzschild spacetime in GR~\cite{Rosa:2011my}. Below, we summarize the application of the matrix continued-fraction method for a generic three-term coupled recurrence relation, following~\cite{Rosa:2011my}. 

The starting point of this method is a matrix three-term coupled recurrence relation: 
\begin{equation}
\label{eq:matrix3termReview}
\mathbf{\Gamma}_1(n)\mathbf{X}_{n+1} + \mathbf{\Gamma}_0(n)\mathbf{X}_n + \mathbf{\Gamma}_{-1}(n)\mathbf{X}_{n-1} = 0,
\end{equation}
where $\mathbf{X}_n$ is a column vector collecting the expansion coefficients of the coupled functions, and $\mathbf{\Gamma}_i(n)$ are square matrices of fixed dimension that depend on $n$ and on the parameters of the ODEs, such as $\omega$. Henceforth, any quantity in boldface represents a matrix or a vector.

Analogous to the Regge-Wheeler case of Sec.~\ref{sec:rwleaver}, we introduce a matrix ratio, such that
\begin{equation}
\label{eq:matrixRn}
\mathbf{X}_{n+1}=\mathbf{R}_n \, \mathbf{X}_n.
\end{equation}
Substituting the above expression and using $\mathbf{X}_{n-1} = \mathbf{R}_{n-1}^{-1}\mathbf{X}_n$ into Eq.~\eqref{eq:matrix3termReview} yields
\begin{equation}
\left[\mathbf{\Gamma}_1(n) \mathbf{R}_n + \mathbf{\Gamma}_0(n) + \mathbf{\Gamma}_{-1}(n)\mathbf{R}_{n-1}^{-1}\right] \mathbf{X}_n = 0.
\end{equation}
Assuming a non-trivial solution and solving for $\mathbf{R}_{n-1}$ gives the nonlinear matrix recurrence
\begin{equation}
\label{eq:matrixRrecursion}
\mathbf{R}_{n-1} = -\left[\mathbf{\Gamma}_0(n) + \mathbf{\Gamma}_1(n)\mathbf{R}_n\right]^{-1} \mathbf{\Gamma}_{-1}(n).
\end{equation}
 Iterating Eq.~\eqref{eq:matrixRrecursion} generates a matrix-valued continued fraction. Formally, one obtains
\begin{align}
\label{eq:matric-cont-frac}
\mathbf{R}_0 = -\bigg\{\mathbf{\Gamma}_0(1) - \mathbf{\Gamma}_1(1)
\bigg[\mathbf{\Gamma}_0(2) - \mathbf{\Gamma}_1(2)\nonumber\\
\left(\mathbf{\Gamma}_0(3) - \cdots \right)^{-1}
\mathbf{\Gamma}_{-1}(3) \bigg]^{-1}
\mathbf{\Gamma}_{-1}(2)\bigg\}^{-1}
\mathbf{\Gamma}_{-1}(1).
\end{align}
Additionally, the $n=0$ relation gives
\begin{equation}
\left[\mathbf{\Gamma}_0(0) + \mathbf{\Gamma}_1(0)\mathbf{R}_0\right]\mathbf{X}_0 = 0.
\end{equation}
For a nontrivial solution, the determinant of the matrix multiplying $\mathbf{X}_0$ must vanish, leading to the equation
\begin{equation}
\label{eq:detcoupledCF}
\det\!\left[\mathbf{\Gamma}_0(0) + \mathbf{\Gamma}_1(0)\mathbf{R}_0\right] = 0.
\end{equation}
Using the solution to the three-term recursion relation obtained with continued fractions of Eq.~\eqref{eq:matric-cont-frac} in the above equation, one finally obtains 
\begin{align}
\label{eq:detcoupledCF-2}
0 &= \det\!\left\{\mathbf{\Gamma}_0(0) - \mathbf{\Gamma}_1(0)
\bigg[\mathbf{\Gamma}_0(1) - \mathbf{\Gamma}_1(1)
\bigg(\mathbf{\Gamma}_0(2) 
\right. 
\nonumber \\
&\left.
- \mathbf{\Gamma}_1(2)
\left(\mathbf{\Gamma}_0(3) - \cdots \right)^{-1}
\mathbf{\Gamma}_{-1}(3) \bigg)^{-1}
\mathbf{\Gamma}_{-1}(2)\bigg]^{-1}
\mathbf{\Gamma}_{-1}(1)
\right\}.
\end{align}
The QNM frequencies are obtained by solving the above equation, after truncating the matrix continued fraction at a sufficiently large truncation order. At the truncation point, a terminal value for the matrix ratio must be prescribed to close the recurrence relation. This is done by setting 
\begin{equation}
    \mathbf{R}_{n_{\rm max}-1} =-\,\mathbf{\Gamma}^{-1}_{0}(n_{\rm max})  \mathbf{\Gamma}_{-1}(n_{\rm max})\, ,
\end{equation}
in the simplest implementation. The truncation order $n_{\rm max}$ is then increased until the roots of Eq.~\eqref{eq:detcoupledCF} become numerically stable. Different families of QNMs, corresponding to the various fields in the coupled system, are found by solving this equation with different initial guesses for the roots.

\section{Reduction Scheme for Single ODEs}
\label{sec:Reductionsingle}

Leaver extended the single continued-fraction method described in Sec.~\ref{sec:rwleaver} to perturbations of a Reissner-Nordstr\"om black hole~\cite{Leaver:1990zz}. In this case, using the ansatz in Eq.~\eqref{eq:ansatzgr} to solve the master equations leads to a four-term recurrence relation. Leaver then reduced this to a three-term recurrence through Gaussian elimination~\cite{Leaver:1990zz}. This technique was later extended to the study of maximally charged black holes~\cite{Onozawa:1995vu}, where a five-term recurrence relation was first reduced to a four-term recurrence relation and then to a three-term one. The reduction procedure and the associated continued-fraction framework have since been used broadly in black hole QNM calculations beyond the original Schwarzschild problem~\cite{Berti:2009kk,Zimmerman:2015trm}. 

In this section, we present a reduction scheme to further exploit the continued-fraction method by reducing a generic $N$-term recurrence relation to three terms. We begin with a linear ODE and perform the steps required to obtain the corresponding recurrence relation. Then, we present the known reduction of a $5\rightarrow3$ term recurrence relation and generalize it to an $N\rightarrow3$ reduction scheme.

We begin by considering a master linear ODE of the form
\begin{equation}
\label{eq:schematicODE}
\hat{\mathcal{D}}\,\Phi(r) = 0,
\end{equation}
where $\hat{\mathcal{D}}$ is a linear differential operator, and $\Phi(r)$ is the master function to be determined. 
In the relevant applications we focus on, the coefficients of the linear operator $\hat{\mathcal{D}}$ are assumed to be sufficiently regular in the region of interest, so that the solution admits a local series expansion about the chosen expansion point. The function $\Phi(r)$ can then be written in the series form
\begin{equation}
\label{eq:diff-ansatz-Phi}
\Phi(r) = \sum_{n=0}^{\infty} x_n\, u(r)^n \, ,
\end{equation}
where $u(r)$ is a suitably chosen analytic function of the independent variable, e.g.\ $u=r$, $u=1/r$, or $u=(r-2M)/r$, and $x_n$ are the expansion coefficients.

We have attempted to keep the presentation fairly generic, but of course, the method below need not generically work for all differential operators $\hat{\cal{D}}$, and master functions $\Phi(r)$. For example, for the problems of interest to us, the master function represents a radiative degree of freedom, so it must satisfy certain ingoing and outgoing boundary conditions. For such problems, one must pull out an asymptotic factor in front of the summand of Eq.~\eqref{eq:diff-ansatz-Phi}, converting it into a Frobenius series (compare Eq.~\eqref{eq:diff-ansatz-Phi} to Eq.~\eqref{eq:ansatzgr}). The reduction method that follows below, however, is independent of these choices.  

We substitute the series ansatz into Eq.~\eqref{eq:schematicODE} and expand the action of the operator $\hat{\mathcal{D}}$ in powers of $u(r)$. This converts the ODE into an algebraic expression involving $u(r)$, $r$, the summation index $n$, and other parameters of the ODE. After collecting terms with equal powers of $u(r)$, one obtains
\begin{equation}
\label{eq:fullntermRR}
    \sum_{n=0}^{\infty} \sum_{i=0}^{N-1} \gamma_i(n)\, x_n\, u(r)^{n+i} = 0 \,.
\end{equation}
Here, $N$ denotes the total number of terms generated by the expansion, and the coefficients $\gamma_i(n)$ are uniquely determined by the ODE and by the choice of $u(r)$. 

Since Eq.~\eqref{eq:fullntermRR} must hold for arbitrary values of $u(r)$ within the domain of convergence of the series, the coefficient of each power of $u(r)$ must vanish independently. To isolate a common power of $u(r)$, one shifts the summation index according to $n \rightarrow n-i$, so that Eq.~\eqref{eq:fullntermRR} can be rewritten as
\begin{equation}
    \sum_{i=0}^{N-1}\sum_{n=i}^{\infty} \gamma_i(n-i)\, x_{n-i}\, u(r)^n = 0 \, .
\end{equation}
Equating the coefficient of each power $u(r)^n$ to zero then yields the $N$-term recurrence relation
\begin{equation}
    \sum_{i=0}^{N-1} \gamma_i(n-i)\, x_{n-i} = 0 \, .
\end{equation}
This $N$-term recurrence relation holds for values of $n$ where all coefficients are well-defined. For indices with $n<i$, the corresponding terms are absent, or equivalently, one may define $x_k=0$ for $k<0$. The recurrence relation then reduces to a lower-order form near the beginning of the series. For convenience, the recurrence relation can be rewritten in the shifted-index form
\begin{equation}
\label{eq:NtermRR}
    \sum_{i=2-N}^{1} \gamma_i(n)\, x_{n+i} = 0 \, ,
\end{equation}

where, by a slight abuse of notation, $\gamma_i(n)$ now denotes the appropriately reindexed recurrence coefficients. We define this equation up to $i=1$ in analogy with the original work~\cite{Leaver:1985ax}. Equation~\eqref{eq:NtermRR} provides the general recurrence relation satisfied by the series coefficients and serves as the starting point for the continued-fraction analysis presented in this section. 

As discussed above, the procedure outlined in the previous paragraphs does not generally yield a three-term ($N=3$) recurrence relation. The number of terms in the recurrence relation is determined by the structure of the differential operator $\hat{\mathcal{D}}$ and the choice of expansion variable $u(r)$. Even second-order ODEs frequently lead to four-term or higher-order recurrence relations~\cite{Leaver:1990zz,Onozawa:1995vu}. While such relations fully encode the original ODE, they cannot be handled directly by the standard continued-fraction method, which requires a three-term recurrence relation. It is therefore necessary to introduce a systematic reduction procedure that converts a generic $N$-term recurrence relation into an equivalent three-term form, thereby allowing the application of the continued-fraction technique.

The reduction procedure we present here is implemented iteratively, eliminating one term at each step. This reduces an $N$-term recurrence relation to an $(N-1)$-term one, then to an $(N-2)$-term one, and so on, until a three-term recurrence relation is obtained. To illustrate the method, we first demonstrate the reduction of a five-term recurrence relation to a three-term recurrence relation, following Ref.~\cite{Onozawa:1995vu}. We then generalize this process to an arbitrary $N$-term recurrence relation.

\subsection{Five To Three Reduction Scheme} \label{sec:5to3}

We now show how to reduce a five-term recurrence relation to a three-term one. For $N=5$, Eq.~\eqref{eq:NtermRR} takes the form
\begin{align}
\label{eq:5termRR}
    \sum_{i=-3}^{1} \gamma_i(n)\, x_{n+i} = 0,
\end{align}
Our first goal is to reduce it to a four-term recurrence relation of the form
\begin{align}
\label{eq:4termRR}
    \sum_{i=-2}^{1} \tilde{\gamma}_i(n)\, x_{n+i} = 0.
\end{align}
To achieve this goal, we search for expressions for the transformed coefficients $\tilde{\gamma}_i$ in terms of the original coefficients $\gamma_i$. Such expressions can be obtained iteratively by eliminating the lowest-order coefficient in Eq.~\eqref{eq:5termRR} at each value of $n$. This process constructs an equivalent four-term recurrence relation and helps identify the pattern that relates $\tilde{\gamma}_i(n)$ to $\gamma_i(n)$.

Let us see explicitly how this can be achieved. First, we note that when $n < 3$, all coefficients $x_{n+i}$ with negative indices vanish in Eq.~\eqref{eq:5termRR}. As a result, in this range of $n$ values, Eq.~\eqref{eq:5termRR} already reduces to a lower-order recurrence relation, and Eq.~\eqref{eq:4termRR} is satisfied by identifying
\begin{equation}
\label{eq:4-3-small-n}
    \tilde{\gamma}_i(n) = \gamma_i(n), \qquad n < 3,\ \forall\, i.
\end{equation}
For $n=2$, Eq.~\eqref{eq:4termRR} becomes
\begin{equation}
\label{eq:4termn2}
    \tilde{\gamma}_1(2)\, x_3 + \tilde{\gamma}_0(2)\, x_2
    + \tilde{\gamma}_{-1}(2)\, x_1 + \tilde{\gamma}_{-2}(2)\, x_0 = 0,
\end{equation}
and solving this equation for the lowest-order coefficient, $x_0$, yields
\begin{equation}
\label{eq:a0n2}
    x_0 =
    -\frac{1}{\tilde{\gamma}_{-2}(2)}
    \left[
        \tilde{\gamma}_1(2)\, x_3
        + \tilde{\gamma}_0(2)\, x_2
        + \tilde{\gamma}_{-1}(2)\, x_1
    \right].
\end{equation}
For $n=3$, the five-term recurrence relation of Eq.~\eqref{eq:5termRR} becomes
\begin{align}
\label{eq:5termn3}
    \gamma_1(3)\, x_4 + \gamma_0(3)\, x_3 + \gamma_{-1}(3)\, x_2\nonumber\\
    + \gamma_{-2}(3)\, x_1 + \gamma_{-3}(3)\, x_0 = 0.
\end{align}
Substituting the expression for $x_0$ in Eq.~\eqref{eq:a0n2} into Eq.~\eqref{eq:5termn3} eliminates the lowest-order coefficient and results in a four-term recurrence relation of the form of Eq.~\eqref{eq:4termRR}. More precisely, Eq.~\eqref{eq:5termn3} becomes
\begin{align}
&0= x_4 \, \gamma_1(3)
+ x_3\left[\gamma_0(3) - \frac{\gamma_{-3}(3)\,\tilde{\gamma}_1(2)}{\tilde{\gamma}_{-2}(2)}\right] \nonumber \\
&\quad
+ x_2\left[\gamma_{-1}(3) - \frac{\gamma_{-3}(3)\,\tilde{\gamma}_0(2)}{\tilde{\gamma}_{-2}(2)}\right]\nonumber \\
&\quad
+ x_1\left[\gamma_{-2}(3) - \frac{\gamma_{-3}(3)\,\tilde{\gamma}_{-1}(2)}{\tilde{\gamma}_{-2}(2)}\right]\,,
\end{align}
and comparing this with Eq.~\eqref{eq:4termRR} for $n=3$,
\begin{equation}
\label{eq:4termn3}
    \tilde{\gamma}_1(3)\, x_4 + \tilde{\gamma}_0(3)\, x_3
    + \tilde{\gamma}_{-1}(3)\, x_2 + \tilde{\gamma}_{-2}(3)\, x_1 = 0,
\end{equation}
allows us to identify
\begin{align}
\label{eq:n=3-case}
   & \tilde{\gamma}_1(3)= \gamma_1(3) \nonumber \\ 
    &\tilde{\gamma}_0(3)=\gamma_0(3) - \frac{\gamma_{-3}(3)\,\tilde{\gamma}_1(2)}{\tilde{\gamma}_{-2}(2)} \nonumber \\ 
    &\tilde{\gamma}_{-1}(3)=\gamma_{-1}(3) - \frac{\gamma_{-3}(3)\,\tilde{\gamma}_0(2)}{\tilde{\gamma}_{-2}(2)} \\
    &\tilde{\gamma}_{-2}(3)=\gamma_{-2}(3) - \frac{\gamma_{-3}(3)\,\tilde{\gamma}_{-1}(2)}{\tilde{\gamma}_{-2}(2)} . \nonumber
\end{align}
We have thus found a way to map the 5-term recursion relation into a 4-term recursion relation for a given $n$, through the identification of $\tilde{\gamma}_i(3)$ as a function of $\gamma_i(3)$ and $\gamma_i(2)$, since $\tilde{\gamma}_i(2) = \gamma_i(2)$ by Eq.~\eqref{eq:4-3-small-n}.

The elimination process can be generalized to any $n \geq 3$ by considering the four-term recurrence at order $n-1$,
\begin{align}
\label{eq:4termnminus1}
\tilde{\gamma}_{-2}(n-1)x_{n-3} &+ \tilde{\gamma}_{-1}(n-1)x_{n-2} \nonumber\\&+ \tilde{\gamma}_{0}(n-1)x_{n-1} + \tilde{\gamma}_{1}(n-1)x_{n} = 0,
\end{align}
which we can solve for the lowest-index coefficient $x_{n-3}$ via
\begin{align}
\label{eq:anminus3}
x_{n-3} &= \frac{-1}{\tilde{\gamma}_{-2}(n-1)}[\tilde{\gamma}_{-1}(n-1)x_{n-2} + \tilde{\gamma}_{0}(n-1)x_{n-1} + \nonumber \\  & \qquad \qquad + \tilde{\gamma}_{1}(n-1)x_{n}].
\end{align}
Substituting Eq.~\eqref{eq:anminus3} into the five-term recurrence of Eq.~\eqref{eq:5termRR} at order $n$,
\begin{align}
\label{eq:5termn}
\gamma_{-3}(n)x_{n-3} + \gamma_{-2}(n)x_{n-2} + \gamma_{-1}(n)x_{n-1}\nonumber\\ + \gamma_{0}(n)x_{n} + \gamma_{1}(n)x_{n+1} = 0,
\end{align}
eliminates the coefficient $x_{n-3}$ and yields
\begin{align}
0 &= \gamma_{1}(n)x_{n+1} \nonumber\\
&\quad + \left[\gamma_{0}(n) - \frac{\gamma_{-3}(n)\tilde{\gamma}_{1}(n-1)}{\tilde{\gamma}_{-2}(n-1)}\right]x_{n} \nonumber\\
&\quad + \left[\gamma_{-1}(n) - \frac{\gamma_{-3}(n)\tilde{\gamma}_{0}(n-1)}{\tilde{\gamma}_{-2}(n-1)}\right]x_{n-1} \\
&\quad + \left[\gamma_{-2}(n) - \frac{\gamma_{-3}(n)\tilde{\gamma}_{-1}(n-1)}{\tilde{\gamma}_{-2}(n-1)}\right]x_{n-2} \, . \nonumber
\end{align}
By comparing this result to Eq.~\eqref{eq:4termRR}, the transformed coefficients $\tilde{\gamma}_i(n)$ are identified as
\begin{align}
\label{eq:5to4}
   & \tilde{\gamma}_1(n)= \gamma_1(n), \nonumber \\ 
    &\tilde{\gamma}_0(n)=\gamma_0(n) - \frac{\gamma_{-3}(n)\,\tilde{\gamma}_1(n-1)}{\tilde{\gamma}_{-2}(n-1)}, \nonumber \\ 
    &\tilde{\gamma}_{-1}(n)=\gamma_{-1}(n) - \frac{\gamma_{-3}(n)\,\tilde{\gamma}_0(n-1)}{\tilde{\gamma}_{-2}(n-1)}, \\
    &\tilde{\gamma}_{-2}(n)=\gamma_{-2}(n) - \frac{\gamma_{-3}(n)\,\tilde{\gamma}_{-1}(n-1)}{\tilde{\gamma}_{-2}(n-1)} . \nonumber
\end{align}
Clearly, this reduces to the mapping in Eq.~\eqref{eq:n=3-case} when $n=3$.
This completes the explicit reduction of the five-term recurrence relation to a four-term recurrence relation at arbitrary order $n\geq3$.

Our next step is to reduce the four-term recurrence relation given in Eq.~\eqref{eq:4termRR} to a three-term recurrence relation of the form
\begin{equation}
\label{eq:3termRR}
    \sum_{i=-1}^{1} \gamma'_i(n)\, x_{n+i} = 0.
\end{equation}
Since this reduction has been discussed extensively in the literature~\cite{Leaver:1990zz}, we present it here only briefly. The relation between $\gamma'_i(n)$ and $\tilde{\gamma}_i(n)$ is obtained in complete analogy with the procedure described above. One solves Eq.~\eqref{eq:3termRR} iteratively for the lowest-order coefficient at each value of $n$ and substitutes the result into Eq.~\eqref{eq:4termRR}. This yields the reduction scheme
\begin{equation}
    \gamma'_i(n) = \tilde{\gamma}_i(n), \qquad n < 2,\ \forall\, i,
\end{equation}
and for $n \geq 2$,
\begin{align}
\label{eq:4to3}
   & \gamma'_1(n)= \tilde{\gamma}_1(n) \nonumber \\ 
    &\gamma'_0(n)=\tilde{\gamma}_0(n) - \frac{\tilde{\gamma}_{-2}(n)\,\gamma'_1(n-1)}{\gamma'_{-1}(n-1)}  \\ 
    &\gamma'_{-1}(n)=\tilde{\gamma}_{-1}(n) - \frac{\tilde{\gamma}_{-2}(n)\,\gamma'_0(n-1)}{\gamma'_{-1}(n-1)} . \nonumber
\end{align}
The full reduction procedure is completed by expressing the final coefficients $\gamma'_i(n)$ in terms of the original coefficients $\gamma_i(n)$, through the successive application of Eqs.~\eqref{eq:5to4} and~\eqref{eq:4to3}.

\subsection{N to Three Reduction Scheme}

The procedure presented above can be generalized to any recursion relation order $N$. The core idea is that, at each reduction step, the coefficients of the new recurrence relation are expressed algebraically in terms of the coefficients of the previous order. To formalize this, we denote by $\gamma^{(k)}_i(n)$ the coefficients of a $k$-term recurrence relation at a given stage of the reduction.

Let the starting relation be the $N$-term recurrence relation given in Eq.~\eqref{eq:NtermRR}, with coefficients $\gamma^{(N)}_i(n)$. The first reduction step produces an $(N-1)$-term recurrence relation
\begin{equation}
\label{eq:Nm1termRR}
    \sum_{i=2-(N-1)}^{1} \gamma^{(N-1)}_{i}(n)\, x_{n+i} = 0.
\end{equation}
We define
\begin{equation}
L := N-2 
\end{equation}
as the smallest value of $n$ where recurrence relation of Eq.~\eqref{eq:NtermRR} differs from Eq.~\eqref{eq:Nm1termRR} (for example, $L=3$ in the five-term
case discussed previously). For $n < L$, terms involving negative-index coefficients vanish, and the two recurrence relations are identical. Consequently, the coefficients satisfy
\begin{equation}
    \gamma^{(N-1)}_i(n) = \gamma^{(N)}_i(n), \qquad n < L,\ \forall\, i.
\end{equation}

For $n \geq L$, the coefficients of the reduced recurrence relation are obtained by iteratively eliminating the lowest-order coefficient. The expressions derived in Eqs.~\eqref{eq:5to4} and~\eqref{eq:4to3} are thereby generalized. The transformation rule takes the form
\begin{align} \label{eq:NtoNm1}
   & \gamma^{(N-1)}_1(n) = \gamma^{(N)}_1(n), \nonumber  \\
   & \gamma^{(N-1)}_i(n)
   = \gamma^{(N)}_i(n)
   - \frac{
        \gamma^{(N)}_{-L}(n)\,
        \gamma^{(N-1)}_{i+1}(n-1)
     }{
        \gamma^{(N-1)}_{-L+1}(n-1)
     }, \quad i < 1 .
\end{align}
As in Sec.~\ref{sec:5to3}, the reduction procedure is applied iteratively to eliminate one term at a time. Repeating Eq.~\eqref{eq:NtoNm1} successively reduces an $(N-1)$-term recurrence relation to an $(N-2)$-term one, then to an $(N-3)$-term one, and so on. This procedure is continued until a three-term recurrence relation is obtained. At each stage of the reduction, the newly defined coefficients depend only on the coefficients from the previous step. 

After $N-3$ such reductions, the original $N$-term recurrence relation of Eq.~\eqref{eq:NtermRR} is transformed into a three-term recurrence relation of the form
\begin{equation}
\label{eq:Nto3RR}
    \gamma^{(3)}_1(n)\, x_{n+1} + \gamma^{(3)}_0(n)\, x_n + \gamma^{(3)}_{-1}(n)\, x_{n-1} = 0.
\end{equation}
The coefficients $\gamma^{(3)}_i(n)$ are obtained by iteratively applying the transformation rule
\begin{align}
\label{eq:generalReduction}
    \gamma^{(k-1)}_1(n) &= \gamma^{(k)}_1(n) \nonumber
    \\
    \gamma^{(k-1)}_i(n) &=
    \gamma^{(k)}_i(n)
    - \frac{
        \gamma^{(k)}_{-L_k}(n)\,
        \gamma^{(k-1)}_{i+1}(n-1)
    }{
        \gamma^{(k-1)}_{-L_k+1}(n-1)
    }\,, \quad i < 1
\end{align}
at the $k$th step (with $k < N-1$), where $L_k = k-2$. For $n < L_k$, the coefficients satisfy
\begin{equation}
    \gamma^{(k-1)}_i(n) = \gamma^{(k)}_i(n),
\end{equation}
in direct analogy with the lower-order examples discussed above. In this manner, each reduction step eliminates the lowest-order term in the $k$-term recurrence relation. The final three-term recurrence relation of Eq.~\eqref{eq:Nto3RR} can finally be analyzed using the continued-fraction method described in Sec.~\ref{sec:rwleaver}.

\section{Reduction Scheme for Coupled ODEs}
\label{sec:Reductioncoupled}

In beyond GR theories, additional radiative degrees of freedom (beyond the metric tensor) can result in a system of coupled ODEs for the perturbation equations. Such a coupled system,  in turn, gives rise to a coupled, matrix recurrence relation containing more than three terms. In such coupled situations, a different reduction scheme is required. The reduction of a coupled, four-term matrix recurrence relation to a three-term one was first presented in~\cite{Pani:2013pma}. In this section, we generalize that construction to the reduction of an arbitrary coupled, $N$-term matrix recurrence relation to a three-term one.

Consider two coupled ODEs of the form
\begin{subequations}
\label{eq:coupledODE}
\begin{align}
\hat{\mathcal{D}}_1 \Psi(r) = \hat{\mathcal{A}}_1 \Phi(r) \\
\hat{\mathcal{D}}_2 \Phi(r) = \hat{\mathcal{A}}_2 \Psi(r)
\end{align}
\end{subequations}
where $\hat{\mathcal{D}}_i$ and $\hat{\mathcal{A}}_i$ are linear differential operators. In the applications of interest, $\hat{\mathcal{D}}_i$ are second-order differential operators, while $\hat{\mathcal{A}}_i$ are at most first-order differential operators. Schematically, we consider operators and coupling functions of the form
\begin{align}
\hat{\mathcal{D}}_i &= f_{D_i}(\partial_r^2, \partial_r, \partial_r^0) \,, \nonumber \\
\hat{\mathcal{A}}_i &= f_{A_i}(\partial_r, \partial_r^0) \,,
\end{align}
and the operators are assumed to be finite and analytic in the domain of interest. The master functions $\Psi(r)$ and $\Phi(r)$ denote unknown functions that we wish to solve for, which play the role of the radiative master functions. 

Let us combine the coupled ODE system into a matrix ODE equation by introducing the vector field
\begin{equation}
\mathbf{\Theta}(r) =
\begin{pmatrix}
\Psi(r) \\
\Phi(r)
\end{pmatrix}
\end{equation}
and the operator matrix
\begin{equation}
\bm{\hat{\mathcal{H}}} =
\begin{pmatrix}
\hat{\mathcal{D}}_1 & -\hat{\mathcal{A}}_1 \\
-\hat{\mathcal{A}}_2 & \hat{\mathcal{D}}_2
\end{pmatrix}.
\end{equation}
The coupled ODE system can then be written in the compact form
\begin{equation}
\label{eq:matrixODE}
\bm{\hat{\mathcal{H}}} \; \mathbf{\Theta}(r) = 0.
\end{equation}

As in the single ODE case, we seek solutions in the form of a series expansion:
\begin{equation}
\label{eq:series-vector}
\mathbf{\Theta}(r) = \sum_{n=0}^{\infty}\mathbf{X}_n\, [u(r)]^n \, ,
\end{equation}
where $u(r)$ is a chosen analytic function and $\mathbf{X}_n$ is a column vector. Its components encode the expansion coefficients of the functions contained in $\mathbf{\Theta}(r)$. We substitute this expansion into Eq.~\eqref{eq:matrixODE} and evaluate the action of the operator $\bm{\hat{\mathcal{H}}}$ on $\mathbf{\Theta}(r)$. As before, we simplified notation here by choosing an expansion of the form of Eq.~\eqref{eq:series-vector}; however, when dealing with ingoing or outgoing boundary conditions, one should pull out an asymptotic factor, thus expanding the solution in a Frobenius series. 

Using the series ansatz of Eq.~\eqref{eq:series-vector} into the system of ODEs in Eq.~\eqref{eq:matrixODE} yields a matrix-valued algebraic expression involving powers of $u(r)$. After collecting terms with equal powers of $u(r)$, one obtains
\begin{equation}
\label{eq:matrixNRR}
\sum_{n=0}^{\infty} \sum_{i=0}^{N-1} \mathbf{\Gamma}_i(n)\,\mathbf{X}_n\, u(r)^{n+i} = 0 \,.
\end{equation}
Here, each $\mathbf{\Gamma}_i(n)$ is a $2\times2$ matrix determined entirely by the structure of the operator $\bm{\hat{\mathcal{H}}}$ and by the choice of $u(r)$. These matrices encode the coupling between neighboring coefficients in the series.

Since Eq.~\eqref{eq:matrixNRR} must hold for all values of $u(r)$ within the domain of convergence of the series, the coefficient of each independent power of $u(r)$ must vanish. To isolate a common power of $u(r)$, one shifts the summation index according to $n \rightarrow n-i$. Equation~\eqref{eq:matrixNRR} can be rewritten as
\begin{equation}
\sum_{i=0}^{N-1}\sum_{n=i}^{\infty} \mathbf{\Gamma}_i(n-i)\,\mathbf{X}_{n-i}\, u(r)^n = 0 \,,
\end{equation}
and equating the coefficient of each power of $u(r)^n$ to zero yields the recurrence relation
\begin{equation}
\sum_{i=0}^{N-1} \mathbf{\Gamma}_i(n-i)\,\mathbf{X}_{n-i} = 0 \, .
\end{equation}

As in the scalar case discussed in Sec.~\ref{sec:Reductionsingle}, the above recurrence relation is well-defined only for values of $n$ such that all coefficients in it exist. For indices $n<i$, the corresponding coefficients are absent, or equivalently, $\mathbf{X}_k=\mathbf{0}$ for $k<0$. This reduces the recurrence relation to a lower-order form near the beginning of the series.

Let us now express the coupled, matrix recurrence relation in a shifted-index form, making the dependence on neighboring coefficient vectors explicit. Starting from Eq.~\eqref{eq:matrixNtermRR}, we relabel the coefficients according to $\mathbf{\Gamma}_i(n-i)\rightarrow \mathbf{\Gamma}_i(n)$ and shift the summation range so that the recurrence is centered on $\mathbf{X}_n$. This yields
\begin{equation}
\label{eq:matrixNtermRR}
\sum_{i=2-N}^{1} \mathbf{\Gamma}_i(n) \,\mathbf{X}_{n+i} = 0,
\end{equation}
which explicitly relates a finite set of neighboring vectors ${\mathbf{X}_{n+i}}$ at each order $n$. As before, this form is valid for indices such that all components of $\mathbf{X}_{n+i}$ exist. 

Equation~\eqref{eq:matrixNtermRR} represents the general $N$-term matrix recurrence relation satisfied by the expansion coefficients of the coupled system. Unlike the single ODE case, this recurrence relation encodes the coupling between the underlying functions through the matrix structure of the coefficients $\mathbf{\Gamma}_i(n)$. This formulation captures, in a compact algebraic form, the full content of the original system of coupled ODEs.

The matrix recurrence relation in Eq.~\eqref{eq:matrixNtermRR} serves as the starting point for the reduction scheme developed below. By systematically reducing this $N$-term matrix relation to an equivalent lower-order one, one ultimately obtains a three-term matrix recurrence relation. This final relation can be analyzed using a matrix-valued continued-fraction method. To illustrate the procedure, we first present the reduction scheme for a five-term matrix recurrence relation and then generalize it to an arbitrary $N$-term recurrence relation.

\subsection{Five to Three Matrix Reduction Scheme}

A coupled, five-term matrix recurrence relation may be written as
\begin{equation}
\label{eq:matrix5termRR}
\sum_{i=-3}^{1} \mathbf{\Gamma}_i(n)\mathbf{X}_{n+i} = 0,
\end{equation}
and the objective is to reduce it to a four-term form
\begin{equation}
\label{eq:matrix4termRR}
\sum_{i=-2}^{1} \tilde{\mathbf{\Gamma}}_i(n)\mathbf{X}_{n+i} = 0.
\end{equation}
Achieving this reduction requires expressing the transformed coefficient matrices $\tilde{\mathbf{\Gamma}}_i(n)$ in terms of the known coefficients $\mathbf{\Gamma}_i(n)$. For $n<3$, Eqs.~\eqref{eq:matrix5termRR} and~\eqref{eq:matrix4termRR} are identical, since the coefficients associated with negative indices vanish. It follows that
\begin{equation}
    \tilde{\mathbf{\Gamma}}_i (n) = \mathbf{\Gamma}_i (n) \qquad n < 3,\, \forall\, i.
\end{equation}

The nontrivial relations arise for $n\geq3$. To determine these, Eq.~\eqref{eq:matrix4termRR} is evaluated explicitly at $n=2$,
\begin{equation}
    \tilde{\mathbf{\Gamma}}_{-2}(2)\mathbf{X}_{0} + \tilde{\mathbf{\Gamma}}_{-1}(2)\mathbf{X}_{1} + \tilde{\mathbf{\Gamma}}_{0}(2)\mathbf{X}_{2} + \tilde{\mathbf{\Gamma}}_{1}(2)\mathbf{X}_{3}= 0,
\end{equation}
which can be solved for the lowest-order coefficient $\mathbf{X}_0$,
\begin{align}
\label{eq:X0n2}
   \mathbf{X}_0 = -\left(\tilde{\mathbf{\Gamma}}_{-2}(2)\right)^{-1}\bigg[\tilde{\mathbf{\Gamma}}_{-1}(2)\mathbf{X}_{1}\nonumber \\+ \tilde{\mathbf{\Gamma}}_{0}(2)\mathbf{X}_{2} + \tilde{\mathbf{\Gamma}}_{1}(2)\mathbf{X}_{3}\bigg].
\end{align}
In writing this expression, it is implicitly assumed that $\left(\tilde{\mathbf{\Gamma}}_{-2}(2)\right)^{-1}\tilde{\mathbf{\Gamma}}_{-2}(2)=\mathbb{I}$, where $\mathbb{I}$ denotes the identity matrix. The five-term recurrence relation in Eq.~\eqref{eq:matrix5termRR} is then evaluated at $n=3$,
\begin{align}
    &\mathbf{\Gamma}_{-3}(3)\mathbf{X}_{0} +\mathbf{\Gamma}_{-2}(3)\mathbf{X}_{1} +\mathbf{\Gamma}_{-1}(3)\mathbf{X}_{2} \nonumber\\ &+ \mathbf{\Gamma}_{0}(3)\mathbf{X}_{3} + \mathbf{\Gamma}_{1}(3)\mathbf{X}_{4} =0,
\end{align}
and the expression for $\mathbf{X}_0$ from Eq.~\eqref{eq:X0n2} is substituted in, yielding
\begin{align}
0&= \mathbf{\Gamma}_1(3)\,\mathbf{X}_4 
+ \left[\mathbf{\Gamma}_0(3) -  \mathbf{\Gamma}_{-3}(3) \left(\tilde{\mathbf{\Gamma}}_{-2}(2)\right)^{-1}\tilde{\mathbf{\Gamma}}_{1}(2) \right]\mathbf{X}_3\nonumber \\
&+ \left[\mathbf{\Gamma}_{-1}(3) -  \mathbf{\Gamma}_{-3}(3) \left(\tilde{\mathbf{\Gamma}}_{-2}(2)\right)^{-1}\tilde{\mathbf{\Gamma}}_{0}(2) \right]\mathbf{X}_2 \\
&+ \left[\mathbf{\Gamma}_{-2}(3) -  \mathbf{\Gamma}_{-3}(3) \left(\tilde{\mathbf{\Gamma}}_{-2}(2)\right)^{-1}\tilde{\mathbf{\Gamma}}_{-1}(2) \right]\mathbf{X}_1.\nonumber
\end{align}
By comparison with Eq.~\eqref{eq:matrix4termRR} evaluated at $n=3$, the transformed coefficients are identified as
\begin{align}
    &\tilde{\mathbf{\Gamma}}_1(3)=\mathbf{\Gamma}_1(3), \nonumber\\
    &\tilde{\mathbf{\Gamma}}_0(3)=\mathbf{\Gamma}_0(3) -  \mathbf{\Gamma}_{-3}(3) \left(\tilde{\mathbf{\Gamma}}_{-2}(2)\right)^{-1}\tilde{\mathbf{\Gamma}}_{1}(2), \nonumber\\
    &\tilde{\mathbf{\Gamma}}_{-1}(3)=\mathbf{\Gamma}_{-1}(3) -  \mathbf{\Gamma}_{-3}(3) \left(\tilde{\mathbf{\Gamma}}_{-2}(2)\right)^{-1}\tilde{\mathbf{\Gamma}}_{0}(2), \\
    &\tilde{\mathbf{\Gamma}}_{-2}(3)= \mathbf{\Gamma}_{-2}(3) -  \mathbf{\Gamma}_{-3}(3) \left(\tilde{\mathbf{\Gamma}}_{-2}(2)\right)^{-1}\tilde{\mathbf{\Gamma}}_{-1}(2).\nonumber
\end{align}

The relations obtained above for $n=3$ illustrate the general elimination procedure. For arbitrary $n\geq3$, the four-term recurrence relation at order $(n-1)$ may be written as
\begin{align}
&\tilde{\mathbf{\Gamma}}_{-2}(n-1)\mathbf{X}_{n-3} + \tilde{\mathbf{\Gamma}}_{-1}(n-1)\mathbf{X}_{n-2} \nonumber\\ &+ \tilde{\mathbf{\Gamma}}_{0}(n-1)\mathbf{X}_{n-1} + \tilde{\mathbf{\Gamma}}_{1}(n-1)\mathbf{X}_{n}=0,
\end{align}
which can be solved for the lowest-index coefficient,
\begin{align}
\mathbf{X}_{n-3} = -\left(\tilde{\mathbf{\Gamma}}_{-2}(n-1)\right)^{-1}\bigg[\tilde{\mathbf{\Gamma}}_{-1}(n-1)\mathbf{X}_{n-2}\nonumber \\
+ \tilde{\mathbf{\Gamma}}_{0}(n-1)\mathbf{X}_{n-1} + \tilde{\mathbf{\Gamma}}_{1}(n-1)\mathbf{X}_{n}\bigg].
\end{align}
Substituting this expression into the five-term recurrence relation Eq.~\eqref{eq:matrix5termRR} evaluated at order $n$,
\begin{align}
&\mathbf{\Gamma}_{-3}(n)\mathbf{X}_{n-3} + \mathbf{\Gamma}_{-2}(n)\mathbf{X}_{n-2} + \mathbf{\Gamma}_{-1}(n)\mathbf{X}_{n-1}\nonumber\\ &+ \mathbf{\Gamma}_{0}(n)\mathbf{X}_{n} + \mathbf{\Gamma}_{1}(n)\mathbf{X}_{n+1}=0,
\end{align}
yields a four-term relation. The resulting relation depends on the coefficients $\{\mathbf{X}_{n-2},\mathbf{X}_{n-1},\mathbf{X}_n,\mathbf{X}_{n+1}\}$. By direct comparison with Eq.~\eqref{eq:matrix4termRR}, the transformed coefficients are found to satisfy
\begin{align}
\label{eq:matrix5to4}
&\tilde{\mathbf{\Gamma}}_{1}(n) = \mathbf{\Gamma}_{1}(n),  \\
&\tilde{\mathbf{\Gamma}}_{0}(n) = \mathbf{\Gamma}_{0}(n) - \mathbf{\Gamma}_{-3}(n)\left(\tilde{\mathbf{\Gamma}}_{-2}(n-1)\right)^{-1}\tilde{\mathbf{\Gamma}}_{1}(n-1),\nonumber \\
&\tilde{\mathbf{\Gamma}}_{-1}(n) = \mathbf{\Gamma}_{-1}(n) - \mathbf{\Gamma}_{-3}(n)\left(\tilde{\mathbf{\Gamma}}_{-2}(n-1)\right)^{-1}\tilde{\mathbf{\Gamma}}_{0}(n-1), \nonumber\\
&\tilde{\mathbf{\Gamma}}_{-2}(n) = \mathbf{\Gamma}_{-2}(n) - \mathbf{\Gamma}_{-3}(n)\left(\tilde{\mathbf{\Gamma}}_{-2}(n-1)\right)^{-1}\tilde{\mathbf{\Gamma}}_{-1}(n-1). \nonumber
\end{align}
These expressions define a recursive mapping between the original five-term coefficients $\mathbf{\Gamma}_i(n)$ and the reduced four-term coefficients $\tilde{\mathbf{\Gamma}}_i(n)$, completing the elimination of the lowest-order term at each step. Throughout this calculation, the matrix $\tilde{\mathbf{\Gamma}}_{-2}(n-1)$ is assumed to be invertible for all the relevant values of $n$. However, no other matrices are required to be invertible for this reduction scheme to work.

The next step is to reduce the problem to a three-term form
\begin{equation}
\label{eq:matrix3termRR}
    \sum_{i=-1}^1 \mathbf{\Gamma}'_i (n)\mathbf{X}_{n+i} = 0. 
\end{equation}
The relation between $\mathbf{\Gamma}'_i$ and $\tilde{\mathbf{\Gamma}}_i$ is obtained in much the same way as shown above. For each value of $n$, one solves Eq.~\eqref{eq:matrix3termRR} for the lowest order coefficient and substitutes it into~\eqref{eq:matrix4termRR}. 
As this reduction has been presented in the literature in the past \cite{Pani:2013pma}, we omit further details here. 

The resulting relations between the reduced coefficients are
\begin{equation}
        \mathbf{\Gamma}'_i (n) = \tilde{\mathbf{\Gamma}}_i (n) \qquad n < 2,\, \forall\, i.
\end{equation}
and for $n\geq 2$:
\begin{align}
\label{eq:matrix4to3}
    &\mathbf{\Gamma}'_1(n) = \tilde{\mathbf{\Gamma}}_1(n), \nonumber \\
    &\mathbf{\Gamma}'_0(n) = \tilde{\mathbf{\Gamma}}_0(n) - \tilde{\mathbf{\Gamma}}_{-2}(n)\left(\mathbf{\Gamma}'_{-1}(n-1)\right)^{-1}\mathbf{\Gamma}'_1(n-1),  \\
    &\mathbf{\Gamma}'_{-1}(n) = \tilde{\mathbf{\Gamma}}_{-1}(n) - \tilde{\mathbf{\Gamma}}_{-2}(n)\left(\mathbf{\Gamma}'_{-1}(n-1)\right)^{-1}\mathbf{\Gamma}'_0(n-1)\nonumber.
\end{align}
These expressions define the recursive mapping between the four-term matrix coefficients $\tilde{\mathbf{\Gamma}}_i(n)$ and the three-term coefficients $\mathbf{\Gamma}'_i(n)$. This thereby completes the reduction to a three-term matrix recurrence relation suitable for the matrix continued-fraction analysis. Again, the new matrix with the lowest index is the only one that is assumed to be invertible.

\subsection{N to Three Matrix Reduction Scheme}

The reduction scheme presented in the previous section can be extended such that it is valid for any starting value of $N$. The core of the procedure is that after each reduction, the new coefficients are expressed in terms of the previous ones. Like the single ODE case, we remind the reader that $\mathbf{\Gamma}_i^{(k)}$ are the coefficients of a $k$th-order recurrence relation. 

The beginning relation is thus an $N$th-term recurrence relation like Eq.~\eqref{eq:matrixNRR} with coefficients $\mathbf{\Gamma}_i^{(N)}$:
\begin{equation}
\label{eq:matrixNtermRRnewnot}
    \sum_{i=2-N}^{1} \mathbf{\Gamma}_i^{(N)}\,\mathbf{X}_{n+i} = 0.
\end{equation}
The first step of the scheme is to reduce this recurrence relation to an $(N-1)$th recurrence relation of the form
\begin{equation}
\label{eq:matrixNm1termRRnewnot}
    \sum_{i=2-(N-1)}^{1} \mathbf{\Gamma}_i^{(N-1)}\, \mathbf{X}_{n+i} = 0.
\end{equation}
To do so, we define $L=N-2$, just as in the scalar case, as the value for which Eqs.~\eqref{eq:matrixNtermRRnewnot} and~\eqref{eq:matrixNm1termRRnewnot} are no longer identical. This is enough for a generalization of the relation between new and old coefficients presented in Eqs.~\eqref{eq:matrix5to4} and~\eqref{eq:matrix4to3}. For $n<L$, the equations are identical and therefore
\begin{align}
    \mathbf{\Gamma}^{(N-1)}_i(n) &= \mathbf{\Gamma}^{(N)}_i(n), \qquad n < L,\ \forall\, i.
\end{align}
When $n\geq L$ and the two equations differ from one another, the reduction takes the form 
\begin{align}
    \mathbf{\Gamma}^{(N-1)}_1(n) &= \mathbf{\Gamma}^{(N)}_1(n),  \\
    \mathbf{\Gamma}^{(N-1)}_i(n) &= \mathbf{\Gamma}^{(N)}_i(n) -\mathbf{\Gamma}^{(N)}_{-L}(n)\nonumber\\&\times\left(\mathbf{\Gamma}^{(N-1)}_{-L+1}(n-1)\right)^{-1}\mathbf{\Gamma}^{(N-1)}_{i+1}(n-1) , \quad i < 1 .\nonumber
\end{align}
In accordance with the procedure outlined in Sec.~\ref{sec:Reductionsingle}, this process is repeated to first obtain an $(N-2)$-term recurrence relation, and then subsequently reduce the order of the recurrence relation, until a three-term matrix recurrence relation is obtained.

After reduction $N-3$ times, the original matrix relation becomes 
\begin{equation}
    \mathbf{\Gamma}^{(3)}_1(n) \mathbf{X}_{n+1} + \mathbf{\Gamma}^{(3)}_0(n) \mathbf{X}_{n} + \mathbf{\Gamma}^{(3)}_{-1}(n) \mathbf{X}_{n-1} =0. 
\end{equation}
 The coefficients $\mathbf{\Gamma}^{(3)}_i(n)$ are obtained by iteratively applying the transformation rule
\begin{align}
\label{eq:matrixgeneralReduction}
    \mathbf{\Gamma}^{(k-1)}_1(n) &= \mathbf{\Gamma}^{(k)}_1(n)\,,
    \\
    \mathbf{\Gamma}^{(k-1)}_i(n) &=
    \mathbf{\Gamma}^{(k)}_i(n)-
     \mathbf{\Gamma}^{(k)}_{-L_k}(n)\nonumber\\&\times
        \left(\mathbf{\Gamma}^{(k-1)}_{-L_k+1}(n-1)\right)^{-1}\,
        \mathbf{\Gamma}^{(k-1)}_{i+1}(n-1),
\end{align}
with $k$ denoting the number of terms after $k$ reductions and $L_k = k-2$. For values of $n < L_k$, the coefficients obey
\begin{equation}
    \mathbf{\Gamma}^{(k-1)}_i(n) = \mathbf{\Gamma}^{(k)}_i(n)\,.
\end{equation}
Similarly to the case presented in the previous subsection, at each step $k$ of this reduction scheme, the matrix $\mathbf{\Gamma}^{(k-1)}_{-L_k+1}(n-1)$ is the only one required to be invertible. 
The three-term matrix recurrence relation encodes all the information of the original recurrence relation. Furthermore, it allows for the use of the matrix continued-fraction method to solve the recurrence relation. 

\section{An example: dynamical Chern--Simons gravity}
\label{sec:dCS}

Now that we have provided the necessary theoretical background for the matrix-valued continued-fraction method for coupled ODEs, we apply it to perturbations of slowly-rotating black holes in dCS gravity. Our goal is to calculate the QNM frequencies of both axial and polar gravitational perturbations. Furthermore, we evaluate the robustness of our method by comparing our frequencies to those obtained via different methods.

\subsection{Field equations in dCS gravity}

We first present a brief introduction to dCS gravity~\cite{Jackiw:2003pm,Alexander:2009tp}. The dCS action in vacuum is
\begin{equation}
    \label{eq:TotalAction}
    S=S_{\rm EH}+S_{\Phi}+S_{\rm CS},
\end{equation}
where 
\begin{subequations}
\label{eqs:action_terms}
\begin{align}
\label{eq:EHAction}
S_{\rm EH} &= \kappa \int d^4x \sqrt{-g}\, R \, , \\
\label{eq:SFAction}
S_{\Phi} &= -\frac{1}{2}\int d^4x \sqrt{-g}\,
g^{\mu\nu}(\nabla_\mu\Phi)(\nabla_\nu\Phi) \, , \\
\label{eq:CSAction}
S_{\rm CS} &= \frac{\alpha}{4}\int d^4x \sqrt{-g}\, \Phi\, {\,^*RR} \, .
\end{align}
\end{subequations}
Here, $g_{\mu\nu}$ is the metric, $g$ is its determinant, $R$ is the Ricci scalar, $\kappa\equiv (16\pi)^{-1}$, $\alpha$ is the dCS coupling constant, $\Phi$ is a massless pseudo-scalar field.
The quantity ${\,^*RR}$ is the Pontryagin density, defined as
\begin{equation}
    ^*RR\equiv {^*R^{\mu}{}_{\nu}}{}^{\kappa\delta}R^{\nu}{}_{\mu\kappa\delta},
\end{equation}
where ${^*R^{\mu}{}_{\nu}}{}^{\kappa\delta}$ is the dual Riemann tensor 
\begin{equation}
    {^*R^{\mu}{}_{\nu}}{}^{\kappa\delta}=\frac{1}{2}\epsilon^\mu{}_{\nu\alpha\beta}R^{\alpha\beta\kappa\delta}\,,
\end{equation}
and $\epsilon^{\mu\nu\alpha\beta}$ is the Levi-Civita tensor.

By varying the action in Eq.~\eqref{eq:TotalAction} with respect to the metric $g_{\mu\nu}$ and the field $\Phi$, one obtains the equations of motion for the gravitational and scalar sectors,
\begin{subequations}
\label{eqs:equations_of_motion}
\begin{align}
\label{eq:graveqmotion}
G_{\mu\nu} + \frac{\alpha}{\kappa} C_{\mu\nu}
&= \frac{1}{2\kappa}\, T^{\Phi}_{\mu\nu} \, , \\
\label{eq:scalareqmotion}
\Box\Phi = - \frac{\alpha}{4}\,{\,^*RR} \, .
\end{align}
\end{subequations}
Here $\Box\equiv\nabla_\mu\nabla^\mu$ is the d'Alembertian operator. The tensors $T^{\Phi}_{\mu\nu}$ and $C_{\mu\nu}$ come from the variation of the actions $S_{\Phi}$ and $S_{\rm CS}$ respectively, and are given by:
\begin{subequations}
\label{eqs:tensors_from_action}
\begin{align}
\label{eq:tensor_T}
T^{\Phi}_{\mu\nu}=(\nabla_\mu\Phi)(\nabla_\nu\Phi)-\frac{1}{2} g_{\mu \nu} (\nabla^\sigma\Phi)(\nabla_\sigma\Phi),\\ 
\label{eq:tensor_C}	
C_{\mu\nu}=(\nabla_\sigma\Phi)\epsilon^{\sigma\delta\alpha}{}_{(\mu}\nabla_\alpha R_{\nu)\delta}+(\nabla_\sigma\nabla_\delta\Phi) {^*R{}^\delta{}_{(\mu\nu)}}{}^\sigma.
\end{align}
\end{subequations}

The field equations naturally contain third derivatives of the metric tensor through the coupling of the metric and the pseudoscalar field. These equations, however, are to be understood perturbatively, because the action is \textit{effective}~\cite{Alexander:2009tp}. This means that the theory and its field equations are only valid within a certain energy cut-off, inside which higher-order curvature terms can no longer be neglected in the action. Working within the cut-off implies working perturbatively away from GR, where dCS effects must be small deformations from GR predictions~\cite{Alexander:2021ssr,Delsate:2014hba,Motohashi:2011pw,Motohashi:2011ds}. This condition naturally avoids all ``causality constraints,'' as recently proven in~\cite{Alexander:2025gdn}. In what follows, we always work within the small-coupling approximation to ensure we stay within the cut-off of the effective theory.     

\subsection{Line element and scalar field for slow rotation}

By solving the field equations  (Eq.~\eqref{eqs:equations_of_motion}) assuming stationarity and axisymmetry, one obtains a slowly-rotating black hole solution in dCS gravity. This solution contains corrections to the Kerr metric due to a pseudoscalar field with primarily dipolar structure (see e.g.~\cite{Alexander:2021ssr}). In this work, we consider the solution derived in~\cite{Yunes:2009hc} and subsequently used, for example, in~\cite{Wagle:2021tam,Wagle:2023fwl} in the context of QNMs. This solution is obtained perturbatively, to linear order in the spin parameter $a$ and to quadratic order in the coupling constant $\alpha$. 

At this order in perturbation theory, the dCS correction modifies the metric linearly, so that the line element can be written as the sum of the slowly-rotating Kerr metric and a dCS correction to the $t\phi$ component:
\begin{equation}
    \label{eq:background-metric}
    ds^2_{\rm dCS}=ds^2_{\rm Kerr}+\frac{5}{4}\frac{\alpha^2}{\kappa}\frac{a}{r^4}\left(1+\frac{12}{7}\frac{M}{r}+\frac{27}{10}\frac{M^2}{r^2}\right)\sin^2{\theta}dtd\phi,
\end{equation}
where the line element of the Kerr metric for slow rotation is
\begin{align}
    ds^2_{\rm Kerr}&=-f(r)dt^2-\frac{4Ma\sin^2{\theta}}{r}dtd\phi+\frac{1}{f(r)}dr^2 \\
    &+r^2d\theta^2+r^2\sin^2{\theta}d\phi^2,
\end{align}
and where recall $f(r)=1-2M/r$ is the Schwarzschild factor. The corresponding pseudoscalar field solution that surrounds this slowly-rotating black hole is
\begin{equation}
    \label{eq:background-scalar}
    \Phi=\frac{5}{8}\frac{a\alpha}{M}\frac{\cos{\theta}}{r^2}\left(1+\frac{2M}{r}+\frac{18M^2}{5r^2}\right)\,.
\end{equation}

The dCS solution above represents a slowly-rotating black hole with ADM mass $M$ and ADM angular momentum $J = Ma$. The dCS correction modifies the frame-dragging sector of the metric and, consequently, the horizon angular frequency. To leading order in spin and coupling constant, this velocity is given by
\begin{equation}
    \Omega_{\rm H}=\frac{a}{4M^2}-\frac{709a\zeta}{28672 M^2}\,,
\end{equation}
where we have defined the dimensionless coupling constant
\begin{align}
    \zeta :=\frac{\alpha^2}{M^4\kappa}.
\end{align}
This modification is important for QNM calculations, as it modifies the asymptotic controlling factor of the metric perturbations.

\subsection{Perturbed Field Equations and the continued-fraction method in dCS gravity}

We are interested in linear perturbations of the slowly-rotating dCS black-hole background introduced above. Accordingly, we perturb the metric and scalar field as
\begin{equation}
\label{eq:linear_perturbations}
{g}_{\mu\nu}=\bar{g}_{\mu\nu}+\epsilon\,\delta g_{\mu\nu},
\qquad
{\Phi}=\bar{\Phi}+\epsilon\,\delta\Phi \, ,
\end{equation}
where $\bar{g}_{\mu\nu}$ and $\bar{\Phi}$ are the background metric and scalar field of Eqs.~\eqref{eq:background-metric} and~\eqref{eq:background-scalar} respectively, $\delta g_{\mu\nu}$ and $\delta\Phi$ are linear perturbations, and $\epsilon$ is a perturbation bookkeeping parameter. 

Substituting Eq.~\eqref{eq:linear_perturbations} into the field equations, expanding to linear order in $\epsilon$, and then working consistently to linear order in $a$ and quadratic order in $\alpha$ (or equivalently, linear order in $\zeta$), one obtains the set of master equations governing the scalar, axial, and polar perturbations. For a detailed explanation of the steps involved, we direct the reader to~\cite{Wagle:2021tam,Srivastava:2021imr,Li:2025fci}. After performing a harmonic decomposition and eliminating the angular dependence, the perturbed field equations can be decoupled into a system of coupled master equations in terms of the scalar master variable $R_{\ell m}(r)$, the axial Regge-Wheeler master function $\Psi^{\rm RW}_{\ell m}(r)$, and the polar Zerilli-Moncrief master function $\Psi^{\rm ZM}_{\ell m}(r)$, namely
\begin{widetext}
\begin{subequations}
\label{eqs:perturbed_final}
\begin{align}
\label{eq:SF}
f(r)^2 \partial_{rr}R_{\ell m}
&+ \frac{2M}{r^2} f(r)\, \partial_r R_{\ell m}
+ \left[\omega^2 - V^S_{\rm eff}(r,a,\alpha^2)\right] R_{\ell m}
=
\alpha f(r)
\left\{
\left[g(r)+a m h(r)\right]\Psi^{\rm RW}_{\ell m}
+ a m j(r)\,\partial_r \Psi^{\rm RW}_{\ell m}
\right\} \, , \\
\label{eq:RW}
f(r)^2 \partial_{rr}\Psi^{\rm RW}_{\ell m}
&+ \frac{2M}{r^2} f(r)\, \partial_r \Psi^{\rm RW}_{\ell m}
+ \left[\omega^2 - V^A_{\rm eff}(r,a,\alpha^2)\right]\Psi^{\rm RW}_{\ell m}
=
\alpha f(r)
\left\{
\left[v(r)+a m n(r)\right] R_{\ell m}
+ a m p(r)\,\partial_r R_{\ell m}
\right\} \, , \\
\label{eq:ZM}
f(r)^2 \partial_{rr}\Psi^{\rm ZM}_{\ell m}
&+ \frac{2M}{r^2} f(r)\, \partial_r \Psi^{\rm ZM}_{\ell m}
+ \left[\omega^2 - V^P_{\rm eff}(r,a,\alpha^2)\right]\Psi^{\rm ZM}_{\ell m}
= 0 \, ,
\end{align}
\end{subequations}   
\end{widetext}
where $m$ is the azimuthal (magnetic) harmonic number, while $[g(r),j(r),v(r),p(r)]$ and the potentials $[V_{\rm{eff}}^S(r),V_{\rm{eff}}^A(r),V_{\rm{eff}}^P(r)]$ are given explicitly in~\cite{Wagle:2021tam}. 
Equation~\eqref{eq:ZM} is decoupled, while Eqs.~\eqref{eq:SF} and~\eqref{eq:RW} form a coupled scalar-axial system. The QNMs of the slowly-rotating dCS black holes are obtained by solving these equations subject to the standard QNM boundary conditions. 

To apply the continued-fraction method to these master equations, one first recasts them in terms of algebraic recurrence relations, as described in Sec.~\ref{sec:Review}. We therefore take the fields
\begin{equation}
\mathbf{\Psi}=\left\{R_{\ell m},\Psi^{\rm RW}_{\ell m},\Psi^{\rm ZM}_{\ell m}\right\}
\end{equation}
to have the form
\begin{equation}
\label{eq:ANZ}
\mathbf{\Psi}(r)=
e^{i\omega r}
(r-r_{\rm H})^{-i\omega_{\rm H} r_{\rm H}}
r^{\,i(2\omega r_{\rm H}-m\Omega_{\rm H} r_{\rm H})}
\sum_{n=0}^{\infty}\mathbf{X}_n\, f(r)^n \, ,
\end{equation}
where $\omega_{\rm H}\equiv \omega-m\Omega_{\rm H}$, $\Omega_{\rm H}$ is the DCS-corrected horizon angular velocity of Eq.~\eqref{eq:background-scalar}, and $\mathbf{X}_n$ denotes the vector of unknown series coefficients. This ansatz is chosen to satisfy the standard QNM boundary conditions, namely,
\begin{equation}
\label{eq:BoundaryConditions}
\mathbf{\Psi}=
\begin{cases}
e^{-i\omega_{\rm H} r_*}, & r\rightarrow r_{\rm H} \, ,\\
e^{i\omega r_*}, & r\rightarrow \infty \, ,
\end{cases}
\end{equation}
where $r_H$ is the location of the horizon given by $r_H = 2M$ for linear-in-$a$, slowly-rotating black holes in dCS gravity.

By substituting the ansatz of Eq.~\eqref{eq:ANZ} into the ODEs of Eq.~\eqref{eqs:perturbed_final} and simplifying, one obtains algebraic recurrence relations for the series coefficients $\mathbf{X}_n$. These relations, like the ODEs themselves, separate into polar and axial sectors. We detail the procedure used to obtain the QNMs in each case in the following two subsections.

Generally, the procedure for obtaining a single QNM is straightforward. One first fixes the mode labels $(\ell,m)$ and chooses a value for the black-hole spin $a$. For each spin, one then varies the coupling constant $\alpha$ and computes the corresponding family of frequencies. Repeating this procedure for different values of the spin produces a grid of QNM frequencies. Since the perturbation equations considered here are valid only to linear order in spin, we restrict attention to slowly-rotating black holes and consider dimensionless spins up to $a/M=0.1$. Likewise, since we work in the small-coupling regime, we restrict to values of the dimensionless coupling up to $\alpha/M^2=0.1$.

In practice, we solve the continued-fraction condition with a complex root finder initialized by the corresponding GR frequencies and then continue the root across the $(a,\alpha^2)$ grid. For the results shown here, we used truncation orders of 
$n_{\rm max} \sim 500 $ for the polar sector and $n_{\rm max} \sim 50 $ for the axial sector, where the scalar field couples to the background spacetime, to achieve relative stability of $10^{-10}$ in both the real and imaginary parts of the frequency. While we could have achieved this 10-digit precision with a smaller number of terms, especially in the polar sector (which does not involve the matrix generalization of the continued-fraction method), we decided to use as many terms as possible in each sector, so as to minimize that source of error. We verified that further increases in $n_{\rm max}$ do not change the results to the quoted digits presented here.

\subsection{Polar gravitational perturbations}

\begin{figure*}[t]
    \centering
    \includegraphics[width=\textwidth]{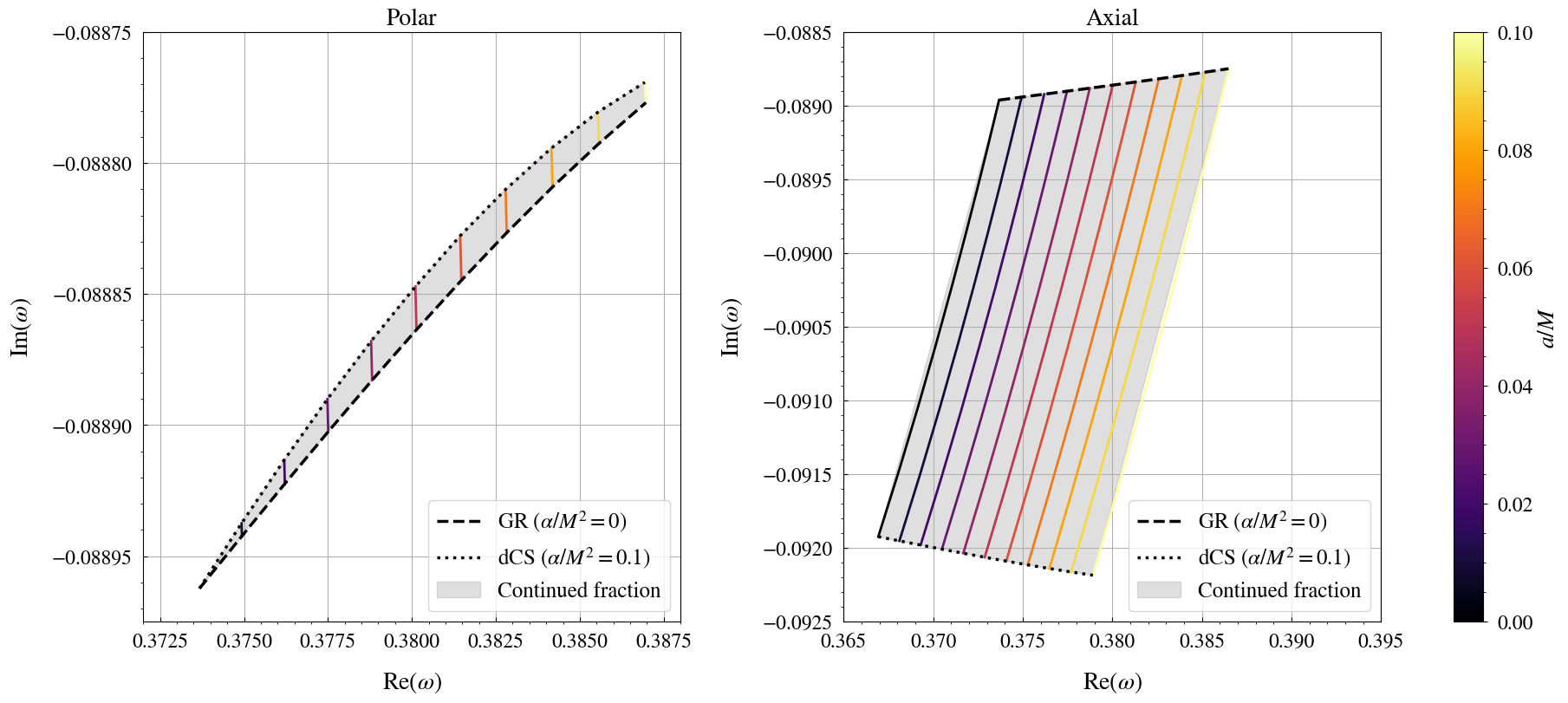}
    \caption{QNM spectrum in dCS gravity obtained with the continued-fraction method. The left panel shows the polar gravitational-led branch, while the right panel shows the axial gravitational-led branch. In each panel, the dashed black curve denotes the GR limit, $\alpha/M^2=0$, and the dotted black curve denotes the dCS boundary at $\alpha/M^2=0.1$. For fixed coupling, varying the spin traces a sequence of QNM frequencies in the complex plane, starting from the GR limit and moving toward the dCS boundary. The colored segments indicate the spin values used in the calculation.}
    \label{fig:Complex_planes}
\end{figure*}

\begin{figure*}[t]
            \centering
             \includegraphics[scale=0.5]{Plots/PolarFracErrors.png}
             \caption{Heatmap of the absolute fractional difference between the QNM frequencies computed with the continued-fraction method and those computed with the MTF-EVP (top row) and METRICS (bottom row) methods, focusing on the real (left column) and imaginary (right column) parts of the $\ell=m=2$ fundamental mode of polar perturbations. The difference in the real or imaginary part never exceeds $10^{-2}$ in both method comparisons, increasing near the $0.1$ limiting values for spin and $\alpha$. }
             \label{fig:Polar_frac_errors}
\end{figure*}

We begin with the polar sector, described by Eq.~\eqref{eq:ZM}. Substituting the ansatz of Eq.~\eqref{eq:ANZ} into this equation yields a single, not-coupled, 16-term recurrence relation. This relation is systematically reduced to a three-term recurrence relation using the single ODE reduction scheme outlined in Sec.~\ref{sec:Reductionsingle}. The coefficients of the initial 16-term recurrence relation are provided in a separate \texttt{Mathematica} notebook \cite{github_repo}. Since the coefficients of the final three-term recurrence relation are very lengthy and uninformative, we do not provide them explicitly here.

The reduced recurrence relation is then analyzed using the continued-fraction method described in Sec.~\ref{sec:Review}. This leads to a condition that is independent of the series coefficients, but that does depend on the mode indices $(\ell,m)$, the black hole spin $a$, the coupling constant $\alpha$, and the complex frequency $\omega$. For fixed values of $(\ell,m)$, $a$, and $\alpha$, and after truncating the continued fraction at a sufficiently large order $N$, one obtains an equation for $\omega$, whose solutions correspond to the fundamental mode and its overtones. In the left panel of Fig.~\ref{fig:Complex_planes}, we present the fundamental frequencies of the dominant $(2,2)$ mode for various spins and coupling constants. The results show that the deviations from GR increase with spin and with the coupling constant, as expected. 
Additional results for the QNMs of the fundamental $\ell=2$ modes with different values of $m$, $\alpha$, and $a$, as well as the first overtones, are provided in Appendix~\ref{appendix:Appendix_A} (Tables~\ref{tab:qnm_polar_fund} and~\ref{tab:qnm_polar_over}); these results are also available to higher precision in~\cite{github_repo}.

While Fig.~\ref{fig:Complex_planes} illustrates the dependence of the QNM frequencies on the parameters $a$ and $\alpha$, QNMs in dCS gravity have been computed using several independent approaches in the literature. To assess the consistency of our results, we perform a quantitative comparison between methods by evaluating the absolute fractional difference between the continued-fraction results and other results, defined as
\begin{subequations}
\label{eqs:fractional_errors}
\begin{align}
\label{eq:real_frac_error}
\delta (\operatorname{Re}(\omega_Y)) &\equiv \left|1 - \frac{\operatorname{Re}(\omega_{\rm continued-fraction})}{\operatorname{Re}(\omega_{Y})} \right|, \\
\label{eq:imag_frac_error}
\delta (\operatorname{Im}(\omega_Y)) &\equiv \left|1 - \frac{\operatorname{Im}(\omega_{\rm continued-fraction})}{\operatorname{Im}(\omega_{Y})} \right|.
\end{align}
\end{subequations}
Here, $Y$ denotes the reference method used for comparison.

In this work, we focus on the fundamental $\ell = m = 2$ mode and compare the continued-fraction results with those obtained using the modified Teukolsky formalism combined with the EVP method (MTF-EVP)~\cite{Li:2025fci}, as well as with results from the metric perturbation framework employing spectral methods (METRICS)~\cite{Chung:2025gyg}. Exact agreement is not expected, as these approaches rely on different approximations. The MTF-EVP method retains terms up to quadratic order in the coupling parameter $\alpha$
and spin, neglecting higher-order contributions, whereas the METRICS approach includes terms quadratic in $\alpha$ but incorporates higher-order corrections in spin. In contrast, our implementation uses a metric background valid to linear order in $a$ and quadratic order in $\alpha$ to obtain the initial recurrence relation; however, during the reduction procedure, the coefficients mix in such a way that higher-order contributions in both $a$ and $\alpha$ are effectively generated in the final three-term recurrence relation. As a result, the QNM frequencies obtained with the continued-fraction method are influenced by these higher-order terms, which are absent in MTF-EVP and only partially included in the METRICS framework.

Figure~\ref{fig:Polar_frac_errors} shows the (base 10) logarithm of the relative fractional difference in the grid created by the range of values for the spin and $\alpha$ that we use to calculate our QNMs. Observe the excellent agreement between the three methods, with the differences never exceeding $10^{-2}$. More specifically, we observe that the error of the continued-fraction method compared to the MTF-EVP one is of ${\cal{O}}(10^{-6})$ to ${\cal{O}}(10^{-2})$ for the real part and ${\cal{O}}(10^{-8})$ to ${\cal{O}}(10^{-3})$ for the imaginary part, with the highest difference occurring when both spin and $\alpha$ are close to 0.1. This is exactly the expected behavior because, as mentioned earlier, the ODEs that we use are linear in spin and quadratic in $\alpha$, and we also study the small rotation regime. As for the difference between continuous-fraction and METRICS results, we see that both the real and imaginary parts agree, with differences in both parts ranging from $10^{-10}$ in the low spin and coupling limit to $10^{-3}$ when we reach our limit of 0.1 for the spin and $\alpha$. 

\subsection{Axial gravitational perturbations}

\begin{figure*}[t]
            \centering
             \includegraphics[scale=0.5]{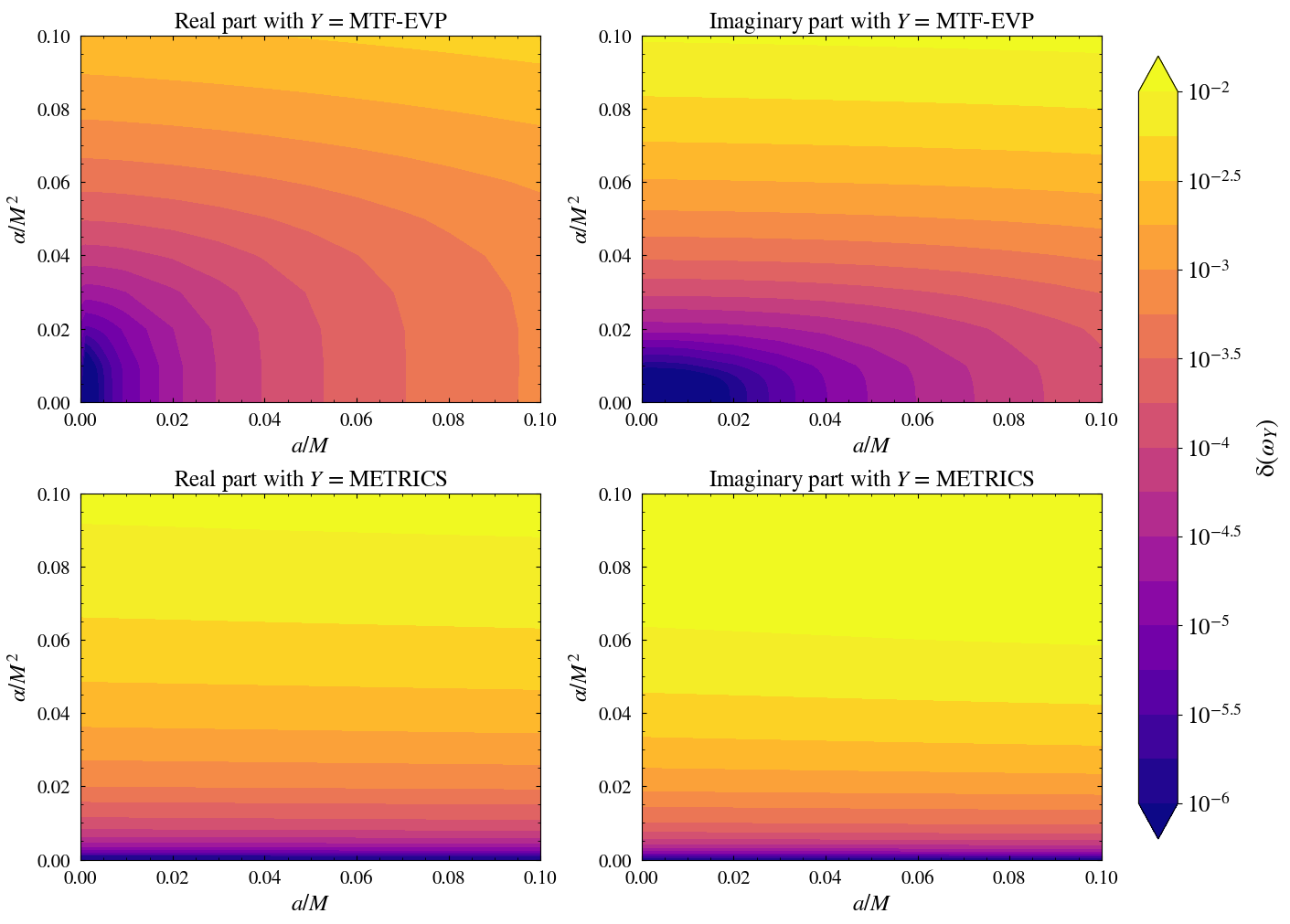}
             \caption{Same as Fig.\ref{fig:Polar_frac_errors} but for axial perturbations. The difference in the real (imaginary) part reaches $10^{-7}$ ($10^{-2}$) and $10^{-2}$ ($2\times10^{-2}$) in the MTF-EVP and METRICS cases, respectively. As in the polar case, the difference increases near the $0.1$ limiting values for spin and $\alpha$. }
             \label{fig:Axial_frac_errors}
\end{figure*} 

In the axial sector, we substitute the ansatz of Eq.~\eqref{eq:ANZ} into Eqs.~\eqref{eq:SF} and~\eqref{eq:RW}, which yields two \textit{coupled}, 12-term recurrence relations. Because the original system is coupled, the series coefficients of the fields \(R_{lm}\) and \(\Psi_{\rm RW}\) appear in both equations. These relations can be recast into a single matrix-valued 12-term recurrence relation with coefficients \(\mathbf{\Gamma}^{(12)}_i(n)\). Among these matrices, the coefficient matrix associated with the lowest index, \(\mathbf{\Gamma}^{(12)}_{-L}(n)\), is non-invertible; however, this poses no difficulty, as the inverse is never required in the reduction procedure. By applying the matrix reduction scheme of Sec.~\ref{sec:Reductioncoupled}, the system is reduced to an equivalent three-term matrix recurrence relation, which can then be solved using a matrix-valued continued-fraction method. The explicit expressions of the reduced coefficients are too lengthy to be presented here, although the original 12-term coefficients are available in a separate \texttt{Mathematica} notebook \cite{github_repo}.

Since the system describes coupled perturbations of the scalar and gravitational fields, the spectrum naturally separates into two families of QNMs, commonly referred to as the scalar-led and gravitational-led sectors. These correspond to distinct sets of roots of the determinant condition of Eq.~\eqref{eq:detcoupledCF} arising in the continued-fraction formulation. In practice, the desired family can be selected by choosing an appropriate initial guess in the root-finding algorithm; for example, using known Schwarzschild black hole QNMs as seeds allows one to isolate the gravitational-led modes. With a suitable choice of initial conditions, both axial gravitational and scalar QNMs can be obtained. As in the polar case, the right panel of Fig.~\ref{fig:Complex_planes} shows clear deviations from the GR predictions for slowly-rotating black holes in dCS gravity. In particular, the imaginary part of the frequencies exhibits a more pronounced deviation than in the polar sector, while the real part shows a comparable shift. Additional results for the $\ell=2$ QNM frequencies at different values of $a$, $\alpha$, and $m$ are presented in the Appendix~\ref{appendix:Appendix_A} for the fundamental mode (Table~\ref{tab:qnm_axial_fund})  and its first overtone (Table~\ref{tab:qnm_axial_over}). Results to higher precision are available in~\cite{github_repo}.

As in the polar case, we assess how consistent our results are with the literature by comparing the QNM frequencies obtained through the continued-fraction method to those obtained with the MTF-EVP and METRICS methods. We compute the relative fractional differences via Eqs.~\eqref{eqs:fractional_errors} and present them in Fig.~\ref{fig:Axial_frac_errors}. Observe that the frequencies agree very well, with differences ranging from $10^{-7}$ to $\sim 10^{-2.5}$ for the real part and from $10^{-7}$ to $10^{-2}$ for the imaginary part. Compared to the polar QNM case, the relative difference is slightly higher here. This is because of higher-order terms in $a$ and $\alpha$ that are generated through the continuous-fraction method. In the axial case, instead of one equation, we now have two coupled ones, which results in a coupled, 12-term matrix recursion relation, whose coefficients become more mixed than in the polar case, via multiple matrix multiplications. As expected, the agreement improves for smaller $a$ and $\alpha$.

\section{Conclusion}
\label{sec:conclusion}

Leaver's continued-fraction method remains one of the most accurate tools for computing black hole QNMs, but in its standard form, it applies only when the Frobenius expansion yields a scalar three-term recurrence relation. Beyond GR, this is generically no longer the case: the recurrence relation may contain more than three terms, may be coupled, or both. In this paper, we showed that these two obstructions can be handled within a single framework, simultaneously. By constructing systematic reduction schemes for arbitrary scalar and matrix $N$-term recurrence relations, we extended the continued-fraction method to a broad class of beyond-GR perturbation problems. This is complementary to recent progress on beyond-GR master equations and modified-Teukolsky formalisms~\cite{Li:2022pcy,Hussain:2022ins,Cano:2023tmv}, which enlarge the class of perturbation problems that can be cast into separable master equations. Our work is also complementary to recent parametrized and generalized continued-fraction implementations in modified-Teukolsky settings~\cite{Cano:2024ezp,Cano:2024jkd}, which apply when the problem remains effectively described by a master equation that is completely decoupled in all degrees of freedom. This paper's contribution addresses the opposite bottleneck: once the perturbation equations are in hand, we show how Leaver-type continued-fraction calculations can still be carried out when the resulting recurrence relation is higher-order \emph{and} coupled. A direct application of this strategy to modified-Teukolsky systems is a natural next step.

As a nontrivial test example, we applied our continued-fraction framework to slowly-rotating black holes in dCS gravity. In this theory, the polar sector yields a 16-term, decoupled scalar recurrence relation, while the axial sector yields a coupled 12-term matrix recurrence relation. After reduction, both systems can be solved with continued fractions. For the fundamental $(\ell,m)=(2,2)$ mode, our frequencies agree well with independent calculations based on the MTF-EVP method and the METRICS spectral framework~\cite{Li:2025fci,Chung:2025gyg}. The largest differences remain at the few-percent level and occur near the edge of the slow-rotation, small-coupling regime studied here, while the agreement is substantially better over much of the parameter space. We also tabulated the remaining $\ell=2$ gravitational-led modes and the first overtone in Appendix~\ref{appendix:Appendix_A}.

The broader payoff of our continued-fraction method is methodological. The framework developed here provides a practical route to accurate continued-fraction calculations whenever beyond-GR perturbation equations produce higher-order or coupled recurrence relations. It therefore opens the door to systematic Leaver-type studies in other theories with non-minimally coupled degrees of freedom and, more importantly, to applications that move beyond the slow-rotation limit when combined with more general perturbation formalisms. If ringdown is to become a precision strong-field test of gravity, we will need methods that are both general and numerically robust, like the one developed here.

\acknowledgements
The authors acknowledge support from the Simons
Foundation through Award No.~896696, the Simons Foundation International through Award No.~SFI-MPS-black hole-00012593-01, the NSF through Grants No.~PHY-2207650 and~PHY-25-12423, and NASA through Grant No. 80NSSC22K0806. P.W. acknowledges funding from the Deutsche Forschungsgemeinschaft (DFG) - project number: 386119226.

\appendix

\onecolumngrid
\section{QNM frequencies for $\ell=2$}
\label{appendix:Appendix_A}

In this appendix, we provide tables for the gravitational QNMs of both axial and polar sectors. The tables include the frequencies for all the multipoles of $\ell=2$ for the fundamental mode $(2,m,0)$ and the first overtone $(2,m,1)$. The data of this paper, along with the series coefficients of the initial recurrence relations, are available online \cite{github_repo}.
\subsection{Polar sector}

The QNM frequencies for the polar sector are shown below. We have evaluated the fundamental (Table~\ref{tab:qnm_polar_fund}) and the first overtone modes for the $\ell=2$ case (Table~\ref{tab:qnm_polar_over}). For each value of the black hole spin, we study three different dCS coupling constant values. Observe that the $m=0$ polar frequencies do not change with $a$ or $\alpha$. This happens because when one sets $m=0$ in Eq.~\eqref{eqs:perturbed_final}, all spin-related terms that are proportional to $a \, m$ vanish, and thus, the master equation reduces to that in GR.  

\begin{table*}[htb]
	\begin{tabular}{c @{\hskip 0.12in} c @{\hskip 0.12in} c @{\hskip 0.12in} c @{\hskip 0.12in} c @{\hskip 0.12in} c @{\hskip 0.12in} c }
		\hline\hline
            $a/M$ & $\alpha/M^2$ & $m=-2$ & $m=-1$ & $m=0$ & $m=1$ & $m=2$ \\
            \hline

            0.00 & 0.00 & .37367, .08896 & .37367, .08896 & .37367, .08896 & .37367, .08896 & .37367, .08896 \\
                & 0.05 & .37367, .08896 & .37367, .08896 & .37367, .08896 & .37367, .08896 & .37367, .08896 \\
                & 0.10 & .37367, .08896 & .37367, .08896 & .37367, .08896 & .37367, .08896 & .37367, .08896 \\
            \hline

            0.01 & 0.00 & .37242, .08898 & .37304, .08897 & .37367, .08896 & .37430, .08895 & .37494, .08894 \\
                & 0.05 & .37242, .08898 & .37305, .08897 & .37367, .08896 & .37430, .08895 & .37493, .08894 \\
                & 0.10 & .37243, .08899 & .37305, .08897 & .37367, .08896 & .37430, .08895 & .37493, .08894 \\
            \hline

            0.02 & 0.00 & .37118, .08900 & .37242, .08898 & .37367, .08896 & .37494, .08894 & .37621, .08892 \\
                & 0.05 & .37119, .08900 & .37242, .08898 & .37367, .08896 & .37493, .08894 & .37621, .08892 \\
                & 0.10 & .37121, .08901 & .37243, .08899 & .37367, .08896 & .37493, .08894 & .37620, .08891 \\
            \hline

            0.03 & 0.00 & .36996, .08902 & .37180, .08899 & .37367, .08896 & .37557, .08893 & .37751, .08890 \\
                & 0.05 & .36997, .08903 & .37180, .08899 & .37367, .08896 & .37557, .08893 & .37750, .08890 \\
                & 0.10 & .36999, .08904 & .37182, .08900 & .37367, .08896 & .37556, .08893 & .37748, .08889 \\
            \hline

            0.04 & 0.00 & .36875, .08904 & .37118, .08900 & .37367, .08896 & .37621, .08892 & .37881, .08888 \\
                & 0.05 & .36876, .08905 & .37119, .08900 & .37367, .08896 & .37621, .08892 & .37881, .08888 \\
                & 0.10 & .36880, .08907 & .37121, .08901 & .37367, .08896 & .37620, .08891 & .37879, .08887 \\
            \hline

            \hline\hline
            \end{tabular}
        \caption{QNM frequencies for the polar gravitational-led sector of the fundamental mode $(2,m,0)$ for slowly-rotating black holes in dCS gravity. The format used is $\operatorname{Re}(\omega),-\operatorname{Im}(\omega)$ with $M=1$, and where the QNM frequencies have been rounded to the 5th decimal digit.}
        \label{tab:qnm_polar_fund}
\end{table*}

\begin{table*}[t]
	\begin{tabular}{c @{\hskip 0.12in} c @{\hskip 0.12in} c @{\hskip 0.12in} c @{\hskip 0.12in} c @{\hskip 0.12in} c @{\hskip 0.12in} c }
		\hline\hline
            $a/M$ & $\alpha/M^2$ & $m=-2$ & $m=-1$ & $m=0$ & $m=1$ & $m=2$ \\
            \hline

            0.00 & 0.00 & .34671, .27392 & .34671, .27391 & .34671, .27391 & .34671, .27391 & .34671, .27391 \\
            & 0.05 & .34671, .27391 & .34671, .27391 & .34671, .27391 & .34671, .27391 & .34671, .27391 \\
            & 0.10 & .34671, .27391 & .34671, .27391 & .34671, .27391 & .34671, .27391 & .34671, .27391 \\
            \hline

            0.01 & 0.00 & .34528, .27404 & .34599, .27398 & .34671, .27391 & .34743, .27385 & .34816, .27379 \\
            & 0.05 & .34527, .27403 & .34599, .27397 & .34671, .27391 & .34744, .27386 & .34817, .27380 \\
            & 0.10 & .34523, .27401 & .34597, .27396 & .34671, .27391 & .34746, .27387 & .34821, .27382 \\
            \hline

            0.02 & 0.00 & .34386, .27417 & .34528, .27404 & .34671, .27391 & .34816, .27379 & .34962, .27366 \\
            & 0.05 & .34384, .27415 & .34527, .27403 & .34671, .27391 & .34817, .27380 & .34964, .27368 \\
            & 0.10 & .34376, .27411 & .34523, .27401 & .34671, .27391 & .34821, .27382 & .34972, .27373 \\
            \hline

            0.03 & 0.00 & .34246, .27430 & .34457, .27411 & .34671, .27391 & .34889, .27372 & .35109, .27353 \\
            & 0.05 & .34242, .27427 & .34455, .27409 & .34671, .27391 & .34891, .27374 & .35113, .27356 \\
            & 0.10 & .34231, .27421 & .34449, .27406 & .34671, .27391 & .34896, .27378 & .35125, .27365 \\
            \hline

            0.04 & 0.00 & .34107, .27442 & .34386, .27417 & .34671, .27391 & .34962, .27366 & .35258, .27341 \\
            & 0.05 & .34102, .27439 & .34384, .27415 & .34671, .27391 & .34964, .27368 & .35264, .27345 \\
            & 0.10 & .34087, .27432 & .34376, .27411 & .34671, .27391 & .34972, .27373 & .35280, .27357 \\
            \hline

            \hline\hline
            \end{tabular}
        \caption{Same table as Table~\ref{tab:qnm_polar_fund}, but for the first overtone $(2,m,1)$.}

     \label{tab:qnm_polar_over}
\end{table*}
\clearpage

\subsection{Axial sector}
Similarly to the polar case, we provide the QNM frequencies for the axial sector for the fundamental mode (Table~\ref{tab:qnm_axial_fund}) and the first overtone modes (Table~\ref{tab:qnm_axial_over}) in the $\ell=2$ case. For each value of the black-hole spin, we once again study 3 different dCS coupling constant values.


\begin{table*}[htb]
	\begin{tabular}{c @{\hskip 0.12in} c @{\hskip 0.12in} c @{\hskip 0.12in} c @{\hskip 0.12in} c @{\hskip 0.12in} c @{\hskip 0.12in} c }
		\hline\hline
            $a/M$ & $\alpha/M^2$ & $m=-2$ & $m=-1$ & $m=0$ & $m=1$ & $m=2$ \\
            \hline

            0.00 & 0.00 & .37367, .08896 & .37367, .08896 & .37367, .08896 & .37367, .08896 & .37367, .08896 \\
                & 0.05 & .37181, .08987 & .37181, .08987 & .37181, .08987 & .37181, .08987 & .37181, .08987 \\
                & 0.10 & .36693, .09192 & .36693, .09192 & .36693, .09192 & .36693, .09192 & .36693, .09192 \\
            \hline

            0.01 & 0.00 & .37242, .08898 & .37304, .08897 & .37367, .08896 & .37430, .08895 & .37493, .08894 \\
                & 0.05 & .37058, .08988 & .37119, .08988 & .37181, .08987 & .37242, .08987 & .37428, .08986 \\
                & 0.10 & .36575, .09189 & .36634, .09191 & .36693, .09192 & .36752, .09194 & .36811, .09195 \\
            \hline

            0.02 & 0.00 & .37117, .08900 & .37242, .08898 & .37367, .08896 & .37493, .08894 & .37620, .08892 \\
                & 0.05 & .36935, .08989 & .37058, .08988 & .37181, .08987 & .37304, .08987 & .37428, .08986 \\
                & 0.10 & .36458, .09187 & .36575, .09189 & .36693, .09192 & .36811, .09195 & .36929, .09198 \\
            \hline

            0.03 & 0.00 & .36992, .08902 & .37179, .08899 & .37367, .08896 & .37556, .08893 & .37747, .08890 \\
                & 0.05 & .36813, .08989 & .36996, .08988 & .37181, .08987 & .37366, .08986 & .37552, .08985 \\
                & 0.10 & .36341, .09184 & .36516, .09188 & .36693, .09192 & .36870, .09197 & .37048, .09201 \\
            \hline

            0.04 & 0.00 & .36868, .08904 & .37117, .08900 & .37367, .08896 & .37620, .08892 & .37874, .08888 \\
                & 0.05 & .36692, .08990 & .36935, .08989 & .37181, .08987 & .37428, .08986 & .37677, .08985 \\
                & 0.10 & .36224, .09181 & .36458, .09187 & .36693, .09192 & .36929, .09198 & .37168, .09204 \\
            \hline

            \hline\hline
            \end{tabular}
        \caption{QNM frequencies for the axial gravitational-led sector of the fundamental mode $(2,m,0)$ for slowly-rotating black holes in dCS gravity. The format used is $\operatorname{Re}(\omega),-\operatorname{Im}(\omega)$ with $M=1$, and where the QNM frequencies have been rounded to the 5th decimal digit.}
        \label{tab:qnm_axial_fund}
\end{table*}

\begin{table*}[h]
	\begin{tabular}{c @{\hskip 0.12in} c @{\hskip 0.12in} c @{\hskip 0.12in} c @{\hskip 0.12in} c @{\hskip 0.12in} c @{\hskip 0.12in} c }
		\hline\hline
            $a/M$ & $\alpha/M^2$ & $m=-2$ & $m=-1$ & $m=0$ & $m=1$ & $m=2$ \\
            \hline

            0.00 & 0.00 & .34671, .27392 & .34671, .27391 & .34671, .27391 & .34671, .27391 & .34671, .27391 \\
            & 0.05 & .34289, .27685 & .34289, .27685 & .34289, .27685 & .34289, .27685 & .34289, .27685 \\
            & 0.10 & .33341, .28345 & .33341, .28345 & .33342, .28344 & .33342, .28345 & .33342, .28345 \\
            \hline

            0.01 & 0.00 & .34527, .27404 & .34599, .27398 & .34671, .27391 & .34743, .27385 & .34815, .27379 \\
            & 0.05 & .34152, .27696 & .34220, .27691 & .34289, .27685 & .34357, .27680 & .34425, .27675 \\
            & 0.10 & .33221, .28345 & .33281, .28345 & .33342, .28345 & .33402, .28345 & .33462, .28345 \\
            \hline

            0.02 & 0.00 & .34384, .27417 & .34527, .27404 & .34671, .27391 & .34815, .27379 & .34960, .27366 \\
            & 0.05 & .34016, .27706 & .34152, .27696 & .34289, .27685 & .34425, .27675 & .34562, .27664 \\
            & 0.10 & .33100, .28345 & .33221, .28345 & .33342, .28345 & .33462, .28345 & .33582, .28345 \\
            \hline

            0.03 & 0.00 & .34241, .27430 & .34456, .27411 & .34671, .27391 & .34887, .27372 & .35104, .27353 \\
            & 0.05 & .33880, .27716 & .34084, .27701 & .34289, .27685 & .34494, .27670 & .34699, .27654 \\
            & 0.10 & .32979, .28344 & .33160, .28345 & .33342, .28345 & .33522, .28345 & .33702, .28344 \\
            \hline

            0.04 & 0.00 & .34098, .27443 & .34384, .27417 & .34671, .27391 & .34960, .27366 & .35249, .27340 \\
            & 0.05 & .33745, .27726 & .34016, .27706 & .34289, .27685 & .34562, .27664 & .34837, .27643 \\
            & 0.10 & .32857, .28343 & .33100, .28345 & .33342, .28345 & .33582, .28345 & .33822, .28343 \\
            \hline

            \hline\hline
            \end{tabular}
        \caption{Same table as Table~\ref{tab:qnm_axial_fund}, but for the first overtone $(2,m,1)$.}
        \label{tab:qnm_axial_over}
\end{table*}

\twocolumngrid

\bibliographystyle{apsrev4-2} 
\bibliography{lib}

\end{document}